\documentclass[letter]{aa}
\usepackage{siunitx}
\usepackage{float}
\usepackage{adjustbox}
\usepackage{adjustbox}
\usepackage{graphicx}	
\usepackage{amsmath}	
\usepackage{amssymb}
\usepackage{hyperref}
\usepackage{tabularx}
\usepackage{mathptmx}
\usepackage{txfonts}
\usepackage{academicons}
\definecolor{orcidlogocol}{HTML}{A6CE39}
\usepackage[T1]{fontenc}
\usepackage{longtable}
\usepackage{subcaption}
\usepackage[symbol]{footmisc}
\usepackage{multicol, blindtext}
\usepackage{subcaption}
\usepackage[font=small]{caption}
\providecommand{\bjdtdb}{\ensuremath{\rm {BJD_{TDB}}}}

\providecommand{\mj}{\ensuremath{\,M_{\rm J}}}
\providecommand{\rj}{\ensuremath{\,R_{\rm J}}}

\providecommand{\fave}{\langle F \rangle} 
\providecommand{\fluxcgs}{10$^9$ erg s$^{-1}$ cm$^{-2}$}

\providecommand{\arcsec}{$^{\prime \prime}$}

\DeclareRobustCommand{\VAN}[3]{#2}
\let\VANthebibliography\thebibliography
\def\thebibliography{\DeclareRobustCommand{\VAN}[3]{##3}\VANthebibliography}
\usepackage{hyperref}
\usepackage{academicons}
\usepackage{xcolor}

\newcommand{\orcid}[1]{\href{https://orcid.org/#1}{\includegraphics[width=8pt]{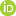}}}

\usepackage{graphicx}
\usepackage{txfonts}
\usepackage{orcidlink}

\begin{document}

   \title{Discovery of a massive giant planet with extreme density around a sub-giant star TOI-4603}

   \author{Akanksha Khandelwal \orcid{0000-0003-0335-6435}
          \inst{1,2}
          \and
          Rishikesh Sharma\inst{1}\orcid{0000-0001-8983-5300} \and Abhijit Chakraborty\inst{1}\orcid{0000-0002-3815-8407} \and Priyanka Chaturvedi\inst{3}\orcid{0000-0002-1887-1192} \and Solène Ulmer-Moll\inst{4,6} \and David~R.~Ciardi\inst{5}\orcid{0000-0002-5741-3047} \and Andrew~W.~Boyle\inst{5}\orcid{0000-0001-6037-2971} \and Sanjay Baliwal\inst{1,2}\orcid{0000-0001-8998-3223} \and Allyson Bieryla\inst{7} \and David W. Latham\inst{7} \and Neelam J.S.S.V. Prasad\inst{1}\orcid{0000-0003-0670-5821} \and Ashirbad Nayak\inst{1} \and Monika Lendl\inst{4} \and Christoph Mordasini\inst{6}
          }

   \institute{Astronomy $\&$ Astrophysics Division, Physical Research Laboratory, Ahmedabad 380009, India\\
              \email{akankshak@prl.res.in}
         \and
             Indian Institute of Technology, 382355 Gandhinagar, India
        \and
            Thüringer Landessternwarte Tautenburg, Sternwarte 5, 07778 Tautenburg, Germany
        \and
            Observatoire de Genève, Université de Genève, Chemin Pegasi, 51, 1290 Versoix, Switzerland
        \and
             NASA Exoplanet Science Institute, Caltech/IPAC, Pasadena, CA 91125, USA
        \and
            Physikalisches Institut, University of Bern, Gesellsschaftstrasse 6, 3012 Bern, Switzerland
        \and
        Center for Astrophysics | Harvard $\&$ Smithsonian, 60 Garden St., Cambridge MA 02138 USA
        }

   \date{Received aaa; accepted bbb}

 
  \abstract
   {We present the discovery of a transiting massive giant planet around TOI-4603, a sub-giant F-type star from NASA's Transiting Exoplanet Survey Satellite (TESS). The newly discovered planet has a radius of $1.042^{+0.038}_{-0.035}$ $R_{J}$, and an orbital period of $7.24599^{+0.00022}_{-0.00021}$ days. Using radial velocity measurements with the PARAS {and TRES} spectrographs, we determined the planet's mass to be $12.89^{+0.58}_{-0.57}$ $M_{J}$, resulting in a bulk density of $14.1^{+1.7}_{-1.6}$ g ${cm^{-3}}$. This makes it one of the few massive giant planets with extreme density and lies in the transition mass region of massive giant planets and low-mass brown dwarfs, an important addition to the population of less than five objects in this mass range. The eccentricity of $0.325\pm0.020$ and an orbital separation of $0.0888\pm0.0010$ AU from its host star suggest that the planet is likely undergoing high eccentricity tidal (HET) migration. We find a fraction of heavy elements of $0.13^{+0.05}_{-0.06}$ and metal enrichment of the planet ($Z_{P}/Z_{star}$) of $4.2^{+1.6}_{-2.0}$. Detection of such systems will offer us to gain valuable insights into the governing mechanisms of massive planets and improve our understanding of their dominant formation and migration mechanisms.}

\keywords{stars: individual: TOI-4603 – planetary systems – techniques: photometric – techniques: radial velocities}

\maketitle
%
\section{Introduction}\label{sec:intro}

Massive giant planets (4 to 13$M_{J}$) have always been debated whether these objects should be classified as planets or Brown dwarfs (BDs)\citep{Chabrier2014,Spiegel2011,Schlaufman2018}. There are a few indirect ways to discriminate between the massive giant planets and the low-mass BDs. One of them is based on the deuterium burning mass-limit, which states that an object should not be massive enough to sustain deuterium fusion at any point in its life to be classified as a planet. The upper mass limit for this deuterium fusion was calculated to be $\approx$13$M_{J}$ for objects of solar metallicity~\citep{boss2005}, regardless of their formation channel. However, the objects with less than 13$M_{J}$ share quite common “nature” with 13$M_{J}$ objects, irrespective of what they have been called. That's why this definition based on a clear-cut mass limit between BDs and planets has caused disagreements~\citep{Chabrier2014}, and various suggestions have been made to reshape it. The study by \citet{Spiegel2011} proposed that deuterium burning may vary from 11 to 16$M_{J}$, depending on the object's helium and other metal content. Some other studies have recommended increasing the upper mass limit to $\sim$ 25$M_{J}$ based on the "driest" region of the brown dwarf desert~\citep{Pont2005,udry2010, Anderson2011}. In another prospect, \citet{artie2015} provided a new definition and preferred the objects in 0.3--60$M_{J}$ to be called Giant Gaseous planets as they follow a particular sequence in the mass-density diagram of all the known planets, sub-stellar objects, and stars (see figure1, \cite{artie2015}). They do not see any abrupt changes in the mass-density diagram for the objects in 0.3-60 $M_{J}$ and suggested irrespective of any formation scenario, these objects should fall under the same general class of objects, i.e., planets. However, recently IAU proposed a working definition of exoplanets\citep{IAU2022} which states that in addition to the 13$M_{J}$ mass limit, the system should have a mass ratio with the central object below the $L_{4}$/$L_{5}$ instability ($M$/$M_{central}$ < 2/(25 + $\sqrt{621}$) $\approx$ 1/25).

Some researchers favor the basis of formation mechanisms to distinguish massive giant planets from BDs. Theoretically, two formation mechanisms dominate the literature: core accretion~\citep[][generally followed by low-mass giant planets ($M_{P}$ < 4$M_{J}$), \citet{Schlaufman2018}]{pollack1996} and Disk instability~\citep[][generally favored by massive giant planets as well as low-mass BDs]{Boss1997}. However, the dominating mechanism for planet formation depends on the disk mass and host star metallicity conditions, i.e., their initial environmental conditions~\citep{Adibekyan2019}. Therefore, it is not clear how to trace the formation history of a planet from the current understanding, and this definition is also inadequate and problematic. Hence, the detailed characterization of more massive giant planets and low-mass BDs will enhance our knowledge of the processes involved in planet formation and offer more insight into the transition regions of these objects.

One frequently debated aspect about the close-in massive giants or giant planets is whether they are formed at their present-day short orbits or migrated from further out orbits~\citep{Batygin2010,baruteau2014}. The common belief is that these planets form beyond the ice line and then migrate inward through various mechanisms to their present location. Migration of a planet to a close-in orbit occurs via torques from the proto-planetary disk (gas disk migration) or gravitational scattering due to another planet or star (HET migration). Eventually, due to tidal forces, its orbit is circularized and shrunk \citep[see][and references therein for a detailed review]{dawson2018}. Nevertheless, recent models suggest the in-situ formation of close-in giant planets is also feasible \citep{Batygin2016} and show that the inner boundary of short-period giant planets and their period-mass distribution could be consistent with predictions for in situ formation \citep{Bailey2018}. That's why which of these three scenarios predominated is still up for discussion, but a combination of these processes likely contributed to the current close-in giant planet population.

In this letter, we report the discovery of TOI-4603 b, a close-in massive giant planet in the overlapping mass-region of BDs and planets. The subsequent sections discuss all the observations, analysis, and results.

\section{Observations}

\subsection{TESS Observations}\label{sec:tess}
\begin{figure}
    \centering
	\includegraphics[width=\columnwidth]{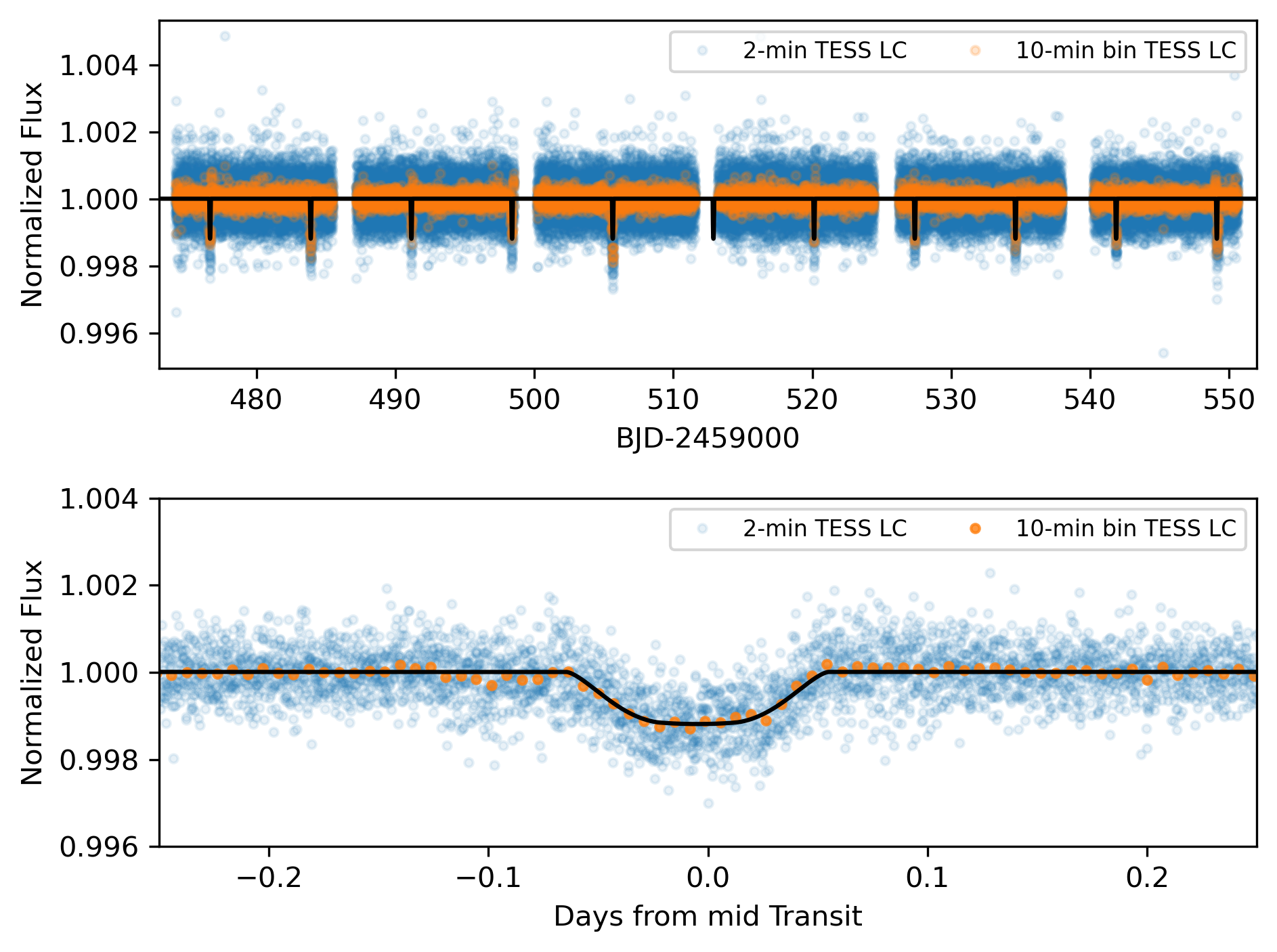}
    \caption {The normalized PDCSAP light curve for TOI-4603 is plotted with respect to time in the upper panel, and time in days from mid-transit is plotted in the lower panel. The 2-min and 10-min binned data points are represented by blue and orange dots, respectively. The black line represents the best-fitted transit model using the EXOFASTv2 (see Section~\ref{sec:global}).}
    \label{fig:tess_lc}
\end{figure}

TESS observed the star TOI-4603 (HD 245134) in three sectors 43, 44, and 45. All the observations were made with the two-minute cadence mode nearly continuously between September 16, 2021, and December 02, 2021 ($\sim 74$ days time-span) with a gap of$~\sim 5.5$ days due to the data transferring from the spacecraft. Light curves were produced and analyzed for transit signals by the Science Processing Operations Center \citep[SPOC:][]{spoc}, consisting of Simple Aperture Photometry (SAP) and Pre-search Data Conditioning Simple Aperture Photometry \citep[PDCSAP:][]{smith_2012,Stumpe_2014} fluxes. These light curves are publicly available at the Mikulski Archive for Space Telescopes (MAST)\footnote{\url{https://mast.stsci.edu/portal/Mashup/Clients/Mast/Portal.html}}. The SPOC pipeline detected ten transits with a depth of $\sim$ 1020 ppm, an orbital period of $\sim$ 7.24 days, and a duration of $\sim$ 2.04 hours. We adopt the median-normalized PDCSAP fluxes for further analysis that are additionally detrended by fitting a high-order polynomial over out-of-transit data using the \texttt{lightkurve} package \citep{lightkurve}. The normalized TESS light curve for TOI-4603 is shown in Figure~\ref{fig:tess_lc}. The Target Pixel Files (TPFs) of TOI-4603 generated with \texttt{tpfplotter} \citep{tpfplot} for all the observing sectors can be found in Figure~\ref{fig:tpfplot}. 

\subsection{High resolution imaging}\label{sec:imaging}
    To assess the possible contamination of bound or unbound close companions on the derived planetary radii \citep{ciardi2015}, we observed the TOI-4603 with near-infrared adaptive optics (AO) imaging at Palomar Observatories. The observations of TOI-4603 were made with the PHARO instrument \citep{hayward2001} behind the natural guide star AO system P3K \citep{dekany2013} on November 21, 2021, in a standard 5-point quincunx dither pattern with steps of 5\arcsec\ in the narrow-band Br$\gamma$ filter $(\lambda_o=2.1686; \Delta\lambda=0.0326~\mu$m). Each dither position was observed three times, offset in position from each other by 0.5\arcsec\ for a total of 15 frames, with an integration time of 5.665 seconds per frame, respectively, for total on-source times of 85 seconds. PHARO has a pixel scale of $0.025\arcsec$ per pixel for a total field of view of $\sim25\arcsec$. The AO data were processed and analyzed with a custom set of IDL tools. The science frames were flat-fielded and sky-subtracted and then combined into a single image using an intra-pixel interpolation that conserves flux, shifts the individual dithered frames by the appropriate fractional pixels, and median-coadds the frames.  The final resolutions of the combined dither were determined from the FWHM of the point spread functions: 0.117\arcsec. The sensitivities of the final combined AO image were determined by injecting simulated sources azimuthally around the primary target every $20^\circ $ at separations of integer multiples of the central source's FWHM \citep{furlan2017}. The brightness of each injected source was scaled until standard aperture photometry detected it with $5\sigma $ significance. The resulting brightness of the injected sources relative to TOI-4603 set the contrast limits at that injection location. The final $5\sigma $ limit at each separation was determined from the average of all of the determined limits at that separation, and the uncertainty on the limit was set by the RMS dispersion of the azimuthal slices at a given radial distance.  The final sensitivity curve for the Palomar data is shown in (Figure~\ref{fig:palomar_ao}); no additional stellar companions were detected.

\subsubsection*{Gaia Assessment}\label{sec:gaia}
    The Gaia Renormalized Unit Weight Error (RUWE) is a metric similar to a reduced chi-square, where values that are $\lesssim 1.4$ indicate that the Gaia astrometric solution is consistent with the star being single, whereas RUWE values $\gtrsim 1.4$ may indicate an astrometric excess noise, possibly caused by the presence of an unseen companion \citep[e.g., ][]{ziegler2020}. TOI-4603 has a Gaia EDR3 RUWE value of 0.998, indicating that the astrometric fits are consistent with the single-star model. 
    
\subsection{Spectroscopy}\label{sec:spec}
\subsubsection{Radial Velocities with PARAS}

RV observations were obtained using the PARAS spectrograph coupled with the 1.2m telescope at PRL Gurushikhar Observatory, Mount Abu, India. PARAS is a fiber-fed echelle spectrograph with a resolving power of R=67000, a wavelength coverage of 380–690~nm. A total of 27 spectra were acquired between January 11, 2022, and November 02, 2022, using the simultaneous wavelength calibration mode with Uranium-Argon (UAr) hollow cathode lamp (HCL) as described in \citet{Chakraborty_2014} and \citet{uar}. The exposure time for all the spectra was 1800 s leading to SNR per pixel of $\sim$  9-18 at the blaze wavelength of 550~nm. More details on observations and data analysis can be found in \citet{Chakraborty_2014}. The reported uncertainties are measured in the same way as described in \cite{10.1093/mnras/stw1560, Chaturvedi_2018, toi1789}. All the RVs and their respective errors are listed in Table~\ref{tab:rv_table}.
\subsubsection{Radial Velocities with TRES}
{We obtained 13 observations between November 3, 2021, and September 16, 2022, using the Tillinghast Reflector Echelle Spectrograph \citep[TRES;][]{gaborthesis} on the 1.5m Tillinghast Reflector telescope on Mount Hopkins, AZ, USA. TRES is a fiber-fed echelle spectrograph with a resolving power of R=44,000 and operating in the wavelength range 390-910 nm. The spectra were obtained in a sequence of 3 observations surrounded by ThAr calibration spectra, and then the median combined to remove cosmic rays. The average exposure time was 290 s resulting in an average SNR per resolution element of 54.2. The spectra were extracted using procedures outlined in \citet{Buchhave2010}, and multi-order relative velocities were derived by cross-correlating the strongest SNR observed spectrum order-by-order against all of the remaining spectra. RVs acquired with TRES spectra with their respective errors are listed in Table~\ref{tab:rv_table}.}

\section{Analysis}
\subsection{Spectroscopic Parameters of TOI-4603}\label{sec:spec_param}
\textcolor{black}{We used the Stellar Parameter Classification \citep[SPC;][]{Buchhave2010,Buchhave2012,Buchhave2014} to derive stellar parameters from TRES spectra. SPC cross-correlates an observed spectrum against a grid of synthetic spectra based on Kurucz atmospheric models \citep{kurucz1992}. Using 12 of the 13 spectra that passed the quality flag based on SNR, we derive $T_\mathrm{eff}$=$6243\pm50$~$K$, $\log{g_*}$ of $3.94\pm0.10$ cgs, [m/H] of $0.22\pm0.08$ dex, and v$\sin{i}$ of $25.70\pm0.50$~km~s$^{-1}$.}

We also obtained high SNR spectra (70 per resolution element at 550~nm) of the 1200 s with Tautenburg coude echelle spectrograph (TCES) installed at the 2m \textcolor{black}{Alfred Jensch telescope, Th\"uringer Landessternwarte Tautenburg, Germany.} TCES is a slit spectrograph with a resolving power of R=67000 and a wavelength coverage of 470–740~nm. For details of the observations, one can refer to \cite{2009A&A...507.1659G}. The spectra were extracted using the IRAF and used to compute stellar parameters with \texttt{zaspe} package \citep{zaspe}. It yields $T_\mathrm{eff}$ of $6273\pm101$~$K$, $\log{g_*}$ of $3.73\pm0.26$ cgs, [Fe/H] of $0.34\pm0.04$ dex, and v$\sin{i}$ of $23.18\pm0.37$~km~s$^{-1}$ through comparison against a grid of synthetic spectra generated from the ATLAS9 model atmospheres \citep{zaspegrid}. \textcolor{black}{The stellar parameters acquired from TRES and TCES spectra are within the error bars.}

Our analysis shows that the TOI-4603 is a metal-rich, F-type sub-giant star. \textcolor{black}{We also calculate the star's rotation period by computing Generalized Lomb-Scargle periodogram \citep[GLS;][]{periodogram} on the out-of-transit TESS PDCSAP light curves and find it to be 5.62 $\pm$ 0.02 days which is comparable to the rotation period (assuming \textit{i}=90) derived using v$\sin{i}$ (section~\ref{sec:spec_param}) and stellar radii (section~\ref{sec:global}). A less significant peak at $\sim$2.28 days was also observed in the periodogram, which may be quasi-periodic and related or unrelated to half of the rotational period signals. Prewhitening the 5.62 days signal did not eliminate the 2.28 days signal, possibly suggesting it originated from another active region on the staller disc. However, further analysis of the 2.28 days signal is beyond the scope of the current work.}

\textcolor{black}{We also inspected the star for solar-like oscillations in the star. We first calculate the expected frequency of the maximum oscillation amplitude ($v_{max}$) using the above calculated $T_\mathrm{eff}$ and $\log{g_*}$ with the seismic scaling relation~\citep{Lund2016}, which yields $v_{max}$ $\approx$700 $\mu$Hz. Since this value is smaller than the Nyquist frequency for the 2-min ($\sim$4166 $\mu$Hz) cadence data, TESS photometric data is well-suited for identifying the oscillations. We analyzed the oscillation signals using the \texttt{lightkurve} package and manually studied the power density spectra of the same TESS light curves but could not detect any significant solar-like oscillations.}

\subsection{Periodogram Analysis}
\textcolor{black}{Independent of photometry, we search for periodic signals in RV data from both spectrographs, PARAS and TRES, using the GLS periodogram. These RVs are corrected for the instrumental offset prior to analysis.} The periodogram is shown in the panel~\texttt{1} of Figure~\ref{fig:periodogram}. Here, we calculate the false alarm probability (FAP) of signals using equations given in \citet{periodogram} and find the most significant signal at 7.24 days (marked with vertical red line in Figure~\ref{fig:periodogram}). This period is the same as estimated from transit data (see Section~\ref{sec:tess}). The signal gives a FAP of 0.007\% at 7.24 days using a bootstrap method over a narrow range centering this period, robustly confirming the periodic signal in our RV data set. The other significant signals in the RV periodogram vanish after removing the 7.24-days periodic signal using a best-fit sinusoidal curve into the datasets (see in panel~\texttt{2}, residual periodogram). The spectral window function is shown in panel~\texttt{3}. As a diagnostics of stellar activity indicator and stellar contamination from nearby stars, we compute the periodogram of bisectors in panel~\texttt{4} and find no statistically significant signal of stellar activity in the data sets.

\subsection{Global modeling}\label{sec:global}
We constrained the system parameters with simultaneous modeling and fitting of \textcolor{black}{the RVs from PARAS and TRES} and the TESS light curves using the publicly available EXOFASTv2 \citep{exofast} package. The software incorporates the differential Evolution Markov Chain Monte Carlo (MCMC) technique with the Bayesian approach to explore all the given parameter space. \\
EXOFASTv2 uses a combination of spectral-energy distribution \citep[SED;][]{SED1} modeling, the stellar evolutionary models; generally MESA Isochrones and Stellar Tracks (MIST) isochrones \citep{mist_choi,mist_dotter}; and the prior parameters to constrain the host star parameters. We performed the SED fitting for TOI-4603 using the broadband photometry from Tycho BV \citep{tycho}, SDSSgri, APASS DR9 BV\citep{APASS}, 2MASS JHK \citep{JHK}, ALL-WISE W1, W2, W3, and W4 \citep{ALLWISE}, as listed in Table~\ref{tab:star_table}.
We imposed Gaussian priors on the $T_\mathrm{eff}$ and [Fe/H], determined from spectral analysis of the TCES spectra. Along with that, we also provided a Gaussian prior on parallax from GaiaDR3 \citep{gaiadr3} and enforced an upper limit on the V-band extinction of 1.59 from \citet{extinction} dust maps at the location of TOI-4603. The SED fitting uses Kurucz’s stellar atmospheric models \citep{Kurucz}, and the resulting best-fitted SED model with broadband photometry fluxes is shown in Figure~\ref{fig:sed}.
Within the EXOFASTv2, the MIST evolutionary tracks are used to provide better estimates for the host star parameters. The most likely MIST evolutionary track from EXOFASTv2 provided the age of $1.64^{+0.30}_{-0.24}$~Gyr (see Figure~\ref{fig:mist}). The adopted stellar parameters are $T_\mathrm{eff}$=$6264^{+95}_{-94}$~K, $\log{g_*}$=$3.810^{+0.021}_{-0.020}$ dex, [Fe/H]=$0.342^{+0.039}_{-0.040}$ dex, $M_{*}$=$1.765\pm0.061$~\(M_\odot\), and $R_{*}$=$2.738^{+0.048}_{-0.050}$~\(R_\odot\). All the parameters are summarized in Table~\ref{result_exofast} along with their 1-$\sigma$ uncertainty.

\begin{figure}
\centering

  \includegraphics[width=\columnwidth]{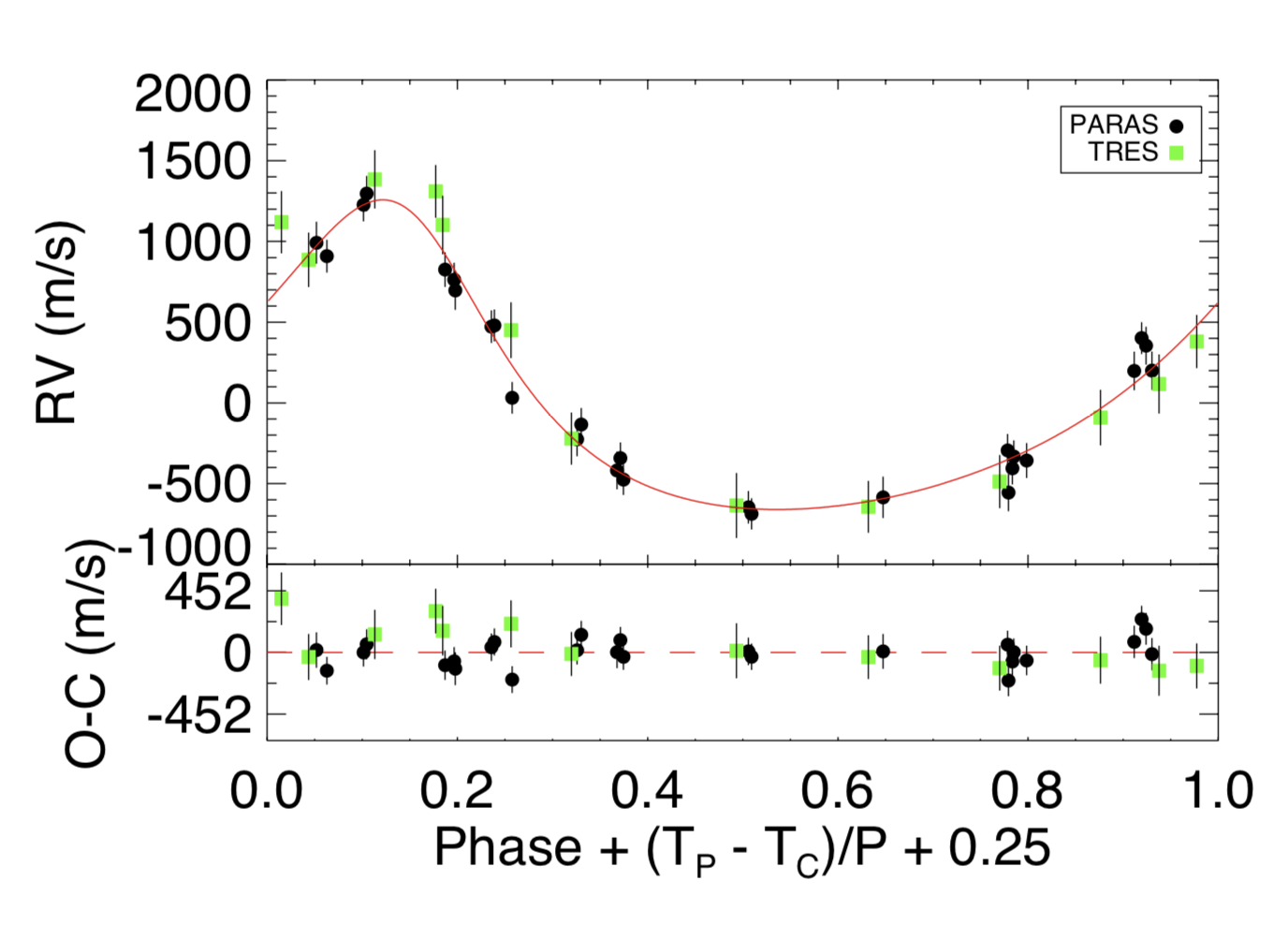}

\caption{The obtained RVs from PARAS and TRES are plotted with respect to $\sim$7.24 days orbital phase. The best-fitted RV model with EXOFASTv2 (see Section~\ref{sec:global}) is represented by the red line, and residuals between the best-fit model and the data are shown in the bottom panel. }
    \label{fig:RV1_curve}
\end{figure}

The simultaneous fitting of the RV and transit data is done by keeping all the parameters (like $b$, $i$, $R_{p}$, $a$ $K$ $\omega$, $e$) free and only providing starting values of $P$ and T$_{c}$ given by the TESS QLP pipeline. The transit model of \citet{Mandel2002} was used for light curve fitting while the RV data is modeled with a standard non-circular Keplerian orbit. We used the default quadratic limb-darkening law for the $TESS$ passband and the limb darkening coefficients ($u_{1}$ and $u_{2}$) were calculated based on tables reported in \citet{Claret} and \citet{Claret_tess}.  We used 42 chains and 50000 steps for each MCMC fit that are further diagnosed for convergence using built-in Gelman-Rubin statistics \citep{Gelman_rubin1992,Gelman_rubin2006}. The transit and RV data with their best-fitted models using EXOFASTv2 are plotted in Figure~\ref{fig:tess_lc}, Figure~\ref{fig:RV1_curve} and Figure~\ref{fig:RV_curve}. In RV data, we also fitted a long-term RV trend ($\dot{\gamma}$) and found it to be $-0.14\pm0.18$ $m s^{-1} day^{-1}$ (Table~\ref{result_exofast}), which may not significant due to its higher uncertainty. All the planetary parameters obtained by EXOFASTv2 are reported in Table~\ref{result_exofast}.

\section{Results and Discussion}

\subsection{TOI-4603 b in context}

\textcolor{black}{We find the mass and radius of TOI-4603 b as $12.89^{+0.58}_{-0.57}$ $M_{J}$ and $1.042^{+0.038}_{-0.035}$ $R_{J}$, respectively, transiting an F-type sub-giant star in a $7.24599^{+0.00022}_{-0.00021}$ days orbit. The discovery of TOI-4603 b is a substantial contribution as it is in the overlapping mass region \citep[11 to 16$M_{J}$;][]{Spiegel2011} of massive giant planets and low-mass brown dwarfs (BDs) based on the deuterium burning mass limit. As per the IAU definition, for solar metallicity, the deuterium burning mass limit is 13$M_{J}$ \citep{IAU2022}. However, this limit depends on other factors, such as the abundance of helium and initial deuterium, and on the metallicity of the invoked model. For example, for three times the solar metallicity, 10$\%$ of initial deuterium can start burning at 11$M_{J}$ \citep{Spiegel2011}. Assuming the metallicity of TOI-4603 b to be the same as that of its parent star, i.e., $0.342^{+0.039}_{-0.040}$ dex, the companion here would have initiated deuterium fusion, loosing on its first criterion to be called a planet. However, according to the second criterion, TOI-4603 b has a mass ratio of 0.007, with the host below the L4/L5 instability ( <1/25), which is in favor of it being called as an exoplanet. Finding the explicit nature of the astrophysical body in this mass range, whether it is a planet or a BD, can be an ambiguous task (see~\citet{Schneider2011} for a detailed overview). For many in the field, including~\citep[][and references therein]{Spiegel2011}, do not consider the deuterium burning mass limit as a strict boundary to distinguish planets and BDs. There are other studies that suggest the upper mass limit for a planet, in particular, a gas-giant planet should be 25$M_{J}$ \citep{Pont2005,udry2010, Anderson2011} and, in some cases 60$M_{J}$\citep{artie2015}. Since TOI-4603 b, according to most of these definitions, qualifies as a gas-giant, we would rather call it a planet in our current work.}

\begin{figure}
	\includegraphics[width=\columnwidth]{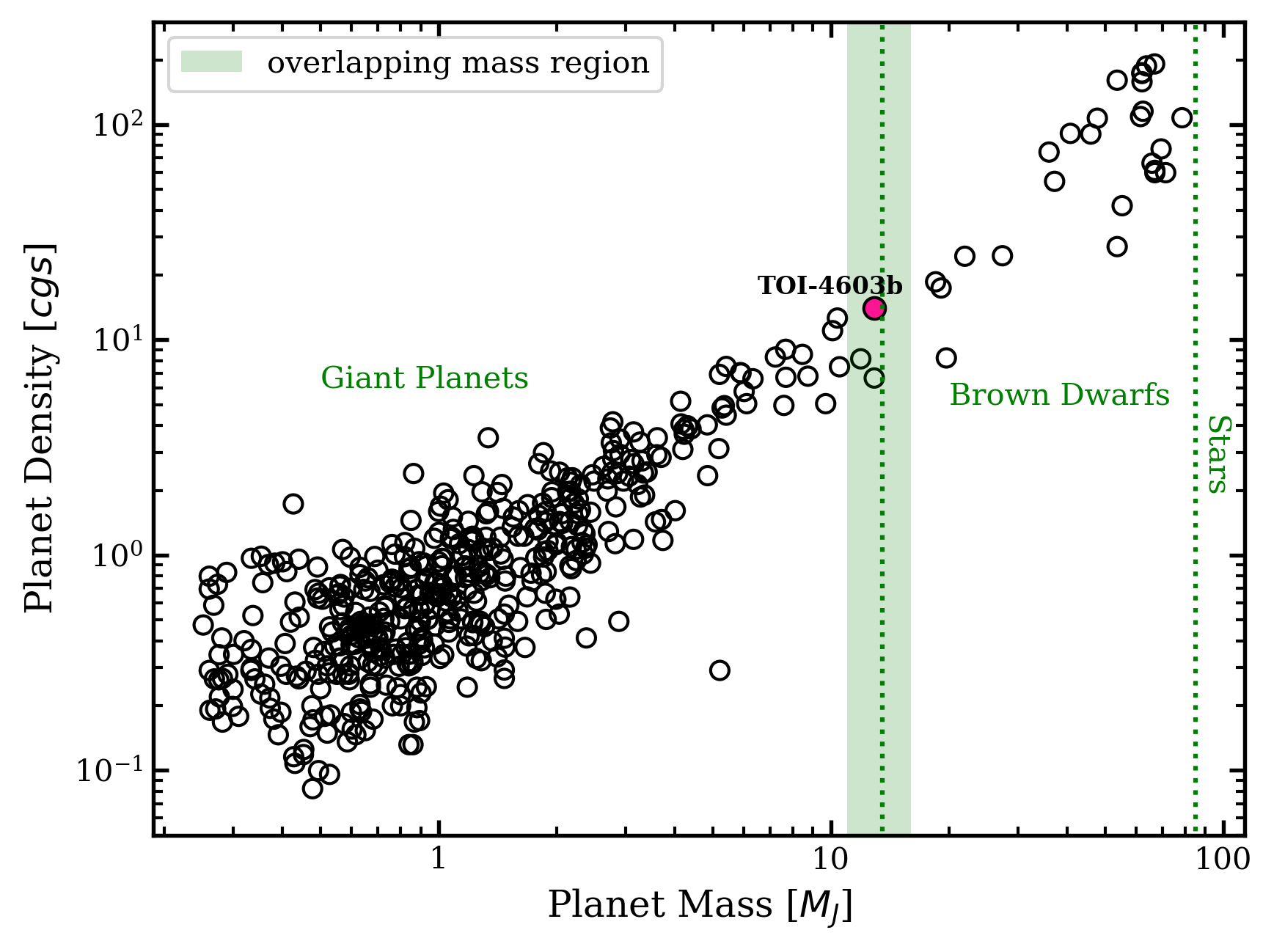}
    \caption{Planetary density as a function of planetary mass for transiting giant planets and brown dwarfs (0.25-85 $M_{J}$). The shaded area represents the overlapping mass region of massive giant planets and brown dwarfs based on the deuterium burning limit, and the dotted lines are at $M_{P}$=13$M_{J}$ and $M_{P}$=85$M_{J}$, respectively. The position of TOI-4603~b is denoted by the magenta dot.}
    \label{fig:sub2}
\end{figure}

\textcolor{black}{We, hereby, present the mass vs. density plot of transiting gas-giant planets and BDs that have mass and radius with a precision better than 25$\%$ with reported mass ranges between 0.25${M_J}$~\citep[lower mass limit for the gas giants from][]{dawson2018} to 85$M_{J}$ (<0.08\(M_\odot\)) in Figure~\ref{fig:sub2}. To date, there are a total of 5310 confirmed exoplanets, out of which 1569 exoplanets’ masses have been determined\footnote{\label{exo_eu}\url{http://exoplanet.eu/}}. Here we focus on the transiting giant planets (0.25-13$M_{J}$), which leave us with 477 transiting giant planets, and of these, 35 are massive giant planets (${M_P}$ > 4${M_J}$)\footnote{\label{tepcat}https://www.astro.keele.ac.uk/jkt/tepcat/ \cite{TEPcat} as of November 16, 2022}. We plot the $M_{P}$=13$M_{J}$, the deuterium fusion mass limit for solar metallicity, as a vertical dotted line and shaded area for planet-BD overlapping mass-region. As seen in the Figure, there have only been three such close-in (a<0.1 AU) transiting objects (HATS-70 b: \citet{hats70} and XO-3 b: \citet{xo3}) discovered in this mass range, including our work. This makes TOI-4603 b an important addition in the context of all the known giant planets.}

\subsection{Internal structure}\label{sec:internal_structure}

We estimate the heavy element content of TOI-4603~b using the method described in \citet{Sarkis2021}. Given the properties of TOI-4603~b, we estimate the planetary radius obtained with the evolution model \texttt{completo21} \citep{Mordasini2012} and compare it with the observed radius. We assume that all the heavy elements are homogeneously mixed in the envelope and are modeled as water with the Equation Of State (EOS) of water ANEOS \citep{Thompson1990, Mordasini2020}. Similarly to \citet{Thorngren2018,Komacek2017}, we do not include a central core. The envelope is coupled with a semi-gray atmospheric model, and hydrogen and helium (He) are modeled with SCvH EOS \citep{Saumon1995} with a He mass fraction Y=0.27.
We use a Bayesian framework to infer the internal luminosity of the planet, which matches the planet's radius given its mass and equilibrium temperature. The internal luminosity is governed by a linear uniform prior, and the content of heavy elements is informed by the \citet{Thorngren2016} relation. We find that the planetary radius is well reproduced with a fraction of heavy elements of $0.13^{+0.05}_{-0.06}$. As noted in \citet{Sarkis2021}, the prior on the internal luminosity has an effect on the final internal luminosity; however, the two values of heavy elements are compatible within $1\,\sigma$. From this fraction of heavy elements in the envelope, we can derive the metal enrichment of the planet $Z_{P}/Z_{star}$=$4.2^{+1.6}_{-2.0}$ (as done in §4.3 from \citet{Ulmer-Moll2022}) and the total mass of heavy elements $M_{z}$=$532^{+205}_{-245}\, M_{\oplus}$. We include in Appendix~D the posterior distribution of the fitted parameters.

TOI-4603b is a scientifically interesting object for studying the processes of planet formation at the transition between massive giant planets and BDs. \citet{Santos2017} proposed two populations of giant planets with masses above and below $\sim$ 4${M_J}$ in their study. Specifically, their finding suggests that the formation of lower-mass giant planets may be related to core accretion and have metal-rich hosts. In contrast, higher-mass planets may form through disk instability mechanisms and orbit stars with lower average metallicity values. Moreover, \citet{Schlaufman2018} established this theory by finding that planets with ${M_P}$ < 4${M_J}$ preferentially orbit metal-rich hosts, unlike planets with ${M_P}$ > 10${M_J}$ do not have this trend. With the high metallicity ([Fe/H]=$0.342^{+0.039}_{-0.040}$ dex), TOI-4603 b does not follow this trend and does not support the existence of any breakout point at 4 $M_{J}$, as suggested by \citet{Adibekyan2019}. It demonstrates that regardless of the metallicity of the host star, a massive giant planet can be formed via any process~\citep{Adibekyan2019}.

\subsection{Eccentricity of TOI-4603~b and tidal circularization}
\begin{figure*}

        \centering
            \includegraphics[width=0.9\textwidth]{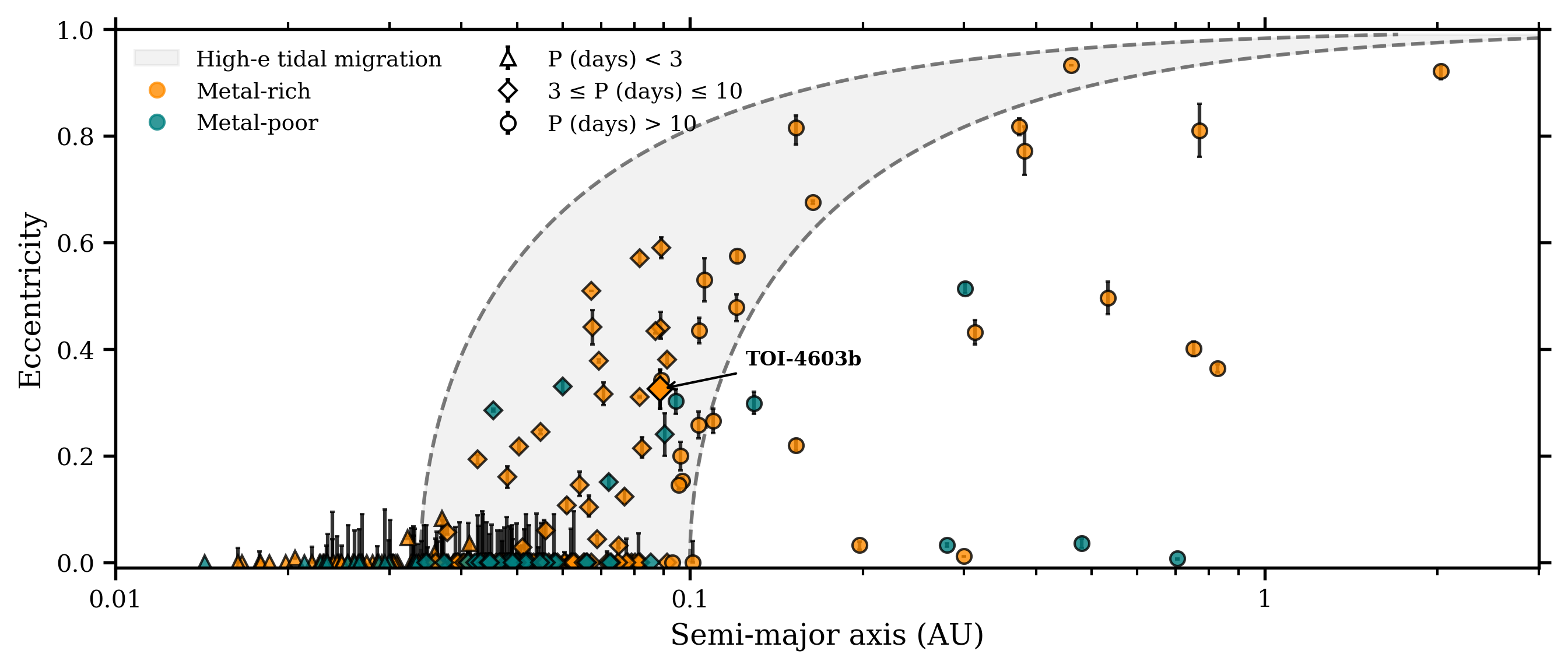}
        
        \caption{The orbital eccentricities of all the transiting giant planets ($0.25M_{J}<M<13M_J$) are plotted with respect to their semi-major axis ($a$) in AU. The datasets are taken from TEPcat database~$^{\ref{tepcat}}$, and planets with a precision better than 25$\%$ in eccentricities are considered. The gray region represents the path of high-eccentricity migration with a range of 0.034-0.1 AU in the final semi-major axis and is set by the Roche limit and the circularization timescale, respectively. The giant planets are color-coded according to their host's metallicity. Triangles represent planets with P<3 days, and diamonds represent planets with 3<P<10 days, and circles represent P> 10 days. The position of TOI-4603~b in the figure is marked with an arrow.}\label{fig:e_vs_a}

\end{figure*}

The orbit of TOI-4603~b is found to be eccentric (e=$0.325\pm0.020$). Different processes, such as secular interactions, planet-planet scattering, planet–disk interactions, and high-eccentricity tidal migration, explain the orbital evolution of giant planets (see Section 2 of \citet{dawson2018} for more details). We plot the observed population of the transiting giant planets (0.25$M_{J}$ < $M_{P}$ < 13$M_{J}$) in eccentricity and semi-major axis parameter space (similar to ~\citet{toi3362b}) in Figure~\ref{fig:e_vs_a} using the TEPcat database$^{\ref{tepcat}}$. The shaded area indicates the region where planets could have undergone HET migration following the constant angular momentum tracks. The boundary of this region is determined by the Roche limit and the tidal circularization timescale, respectively, and is defined as a=0.034–0.1 AU. The position of TOI-4603~b indicates that its orbit is undergoing HET migration. 

Based on the giant planets' eccentricity distribution (Figure~\ref{fig:e_vs_a}), planets with orbital periods between 3 and 10 days have a wider range of eccentricities (0.2 < $e$ < 0.6) than those with shorter periods ($e$ < 0.2). HET migration is the most favorable explanation for these moderate eccentricities, implying that these eccentric giant planets are in the process of tidal circularization. We also observe circular and eccentric giants at the same orbital periods, reason being, circular giant planets started their migration earlier than eccentric giant planets or have more efficient tidal dissipation effects. Some low eccentricities may be due to other formation channels like in situ formation or disk migration.
Furthermore, mostly all eccentric giant planets orbit metal-rich stars, whereas circular giant planets orbit both metal-poor and metal-rich stars (Figure~\ref{fig:e_vs_a}). Given the well-known correlation between the occurrence of giant planets and stellar metallicity, \citet{Dawson-Murray-Clay-2013} established that eccentric giant planets primarily orbit metal-rich stars. Their findings support HET migration via planet-planet gravitational interaction. Being a metallic host, and eccentric orbit of TOI-4603~b is consistent with this trend. \textcolor{black}{Moreover, \citet{gaia_proper_motion} found that the TOI-4603 has a widely separated ($\sim$1.8 AU) BD companion ($M_{P}$ $\approx$ 20.52$M_{J}$) in its orbit. This BD companion of the TOI-4603 system may provide an explanation for this eccentricity.} We also calculated the shortest tidal circularization timescale ($\tau_{cir}$) of 8.2 Gyr~\citep[for Q=10$^5$;][]{circ_time}, greater than the star's current age determined from this work. So, as per tidal evolutionary theory, the orbit of TOI-4603~b has not been circularized, which is consistent with our observations.

\section{Summary and Future prospects}
This work presents the discovery and characterization of a transiting giant planet around a subgiant star TOI-4603 at an orbital period of $7.24599^{+0.00022}_{-0.00021}$ days and was initially identified as an exoplanet candidate using transit observations by NASA's TESS mission. Further, we complemented the TESS data with the ground-based observations with PARAS/PRL, TCES/TLS, TRES, and PHARO/Palomar instruments.
Based on the global modeling of the TOI-4603 system, the host star is found to be metal-rich ([Fe/H]=$0.342^{+0.039}_{-0.040}$ dex), subgiant ($\log{g_{*}}$=$3.810^{+0.021}_{-0.020}$ g ${cm^{-3}}$), F-type ($T_{\rm eff}$=$6264^{+95}_{-94}$ K) star that has a mass, radius, and age of $1.765\pm0.061$ \(M_\odot\), $2.738^{+0.048}_{-0.050}$ \(R_\odot\), and $1.64^{+0.30}_{-0.24}$ Gyr, respectively. The planet TOI-4603~b has a mass of $12.89^{+0.58}_{-0.57}$ $M_{J}$, a radius of $1.042^{+0.038}_{-0.035}$ $R_{J}$, and eccentricity of  $0.325\pm0.020$ with an equilibrium temperature of $1677\pm24$ K. It is one of the most massive and densest transiting giant planets known to date and a valuable addition to the population of less than five massive close-in giant planets in the high-mass planet and low-mass brown dwarf overlapping region (11$M_{J}$ <$M_{P}$ < 16$M_{J}$) that is further required for understanding the processes responsible for their formation. 

TOI-4603 is a rapid rotator ($v\sin{i}$=$23.18\pm0.37$~km~s$^{-1}$) and relatively bright star ($V$=9.2), well suited for Rossiter-McLaughlin effect \citep[R-M;][]{1924ApJ....60...15R, 1924ApJ....60...22M} study and helpful in measuring the projected stellar obliquity of planets. The calculated RM semi-amplitude \citep{rm} for the projected spin-orbit angle ($\lambda$) between $\ang{0}$ and $\ang{90}$ is 6.4~m~s$^{-1}$ and 31~m~s$^{-1}$, respectively. The detection of the RM effect for TOI-4603 is possible by observing precise RVs using moderate-sized telescopes (2.5-4 m aperture); for example, PARAS-2~\citep{paras2} at 2.5~m telescope, PRL is well suited for this work.

\begin{acknowledgements}
      We acknowledge the generous support from PRL-DOS (Department of Space, Government of India) and the director of PRL for the PARAS spectrograph funding for the exoplanet discovery project and research grant for AK and SB. PC acknowledges the generous support from Deutsche Forschungsgemeinschaft (DFG) of the grant HA3279/11-1. We acknowledge the help from Kapil Kumar, Vishal Shah, and all the Mount-Abu, TLS, and Palomar observatory staff for their assistance during the observations. This work was also supported by the Thüringer Ministerium f\"ur Wirtschaft, Wissenschaft und Digitale Gesellschaft. This work has been carried out within the framework of the NCCR PlanetS supported by the Swiss National Science Foundation under grants 51NF40$\_$182901 and 51NF40$\_$205606. We generously acknowledge Dr. Rafael Brahm for providing the grids to determine the spectroscopic parameters using \texttt{zaspe}. PC generously acknowledges Dr. Eike W. Guenther for his contribution in the spectroscopic observations from TCES. This research has made use of the SIMBAD database and the VizieR catalogue access tool, operated at CDS, Strasbourg, France. This research has made use of the Exoplanet Follow-up Observation Program (ExoFOP; DOI: 10.26134/ExoFOP5) website, which is operated by the California Institute of Technology, under contract with the National Aeronautics and Space Administration under the Exoplanet Exploration Program. This paper includes data collected with the TESS mission, obtained from the MAST data archive at the Space Telescope Science Institute (STScI). This work has made use of the Transiting ExoPlanet catalogue (TEPcat) database. We would like to thank the anonymous referee for his/her numerous good suggestions which improved the quality of the paper. 
\end{acknowledgements}
\bibliographystyle{aa}
\bibliography{main}

\begin{thebibliography}{83}
\expandafter\ifx\csname natexlab\endcsname\relax\def\natexlab#1{#1}\fi

\bibitem[{{Adams} \& {Laughlin}(2006)}]{circ_time}
{Adams}, F.~C. \& {Laughlin}, G. 2006, \apj, 649, 1004

\bibitem[{{Adibekyan}(2019)}]{Adibekyan2019}
{Adibekyan}, V. 2019, Geosciences, 9, 105

\bibitem[{{Aller} {et~al.}(2020){Aller}, {Lillo-Box}, {Jones}, {Miranda}, \&
  {Barcel{\'o} Forteza}}]{tpfplot}
{Aller}, A., {Lillo-Box}, J., {Jones}, D., {Miranda}, L.~F., \& {Barcel{\'o}
  Forteza}, S. 2020, \aap, 635, A128

\bibitem[{{Anderson} {et~al.}(2011){Anderson}, {Collier Cameron}, {Hellier},
  {Lendl}, {Maxted}, {Pollacco}, {Queloz}, {Smalley}, {Smith}, {Todd},
  {Triaud}, {West}, {Barros}, {Enoch}, {Gillon}, {Lister}, {Pepe},
  {S{\'e}gransan}, {Street}, \& {Udry}}]{Anderson2011}
{Anderson}, D.~R., {Collier Cameron}, A., {Hellier}, C., {et~al.} 2011, \apjl,
  726, L19

\bibitem[{{Bailey} \& {Batygin}(2018)}]{Bailey2018}
{Bailey}, E. \& {Batygin}, K. 2018, \apjl, 866, L2

\bibitem[{{Baruteau} {et~al.}(2014){Baruteau}, {Crida}, {Paardekooper},
  {Masset}, {Guilet}, {Bitsch}, {Nelson}, {Kley}, \&
  {Papaloizou}}]{baruteau2014}
{Baruteau}, C., {Crida}, A., {Paardekooper}, S.~J., {et~al.} 2014, in
  Protostars and Planets VI, ed. H.~{Beuther}, R.~S. {Klessen}, C.~P.
  {Dullemond}, \& T.~{Henning}, 667

\bibitem[{{Batygin} {et~al.}(2016){Batygin}, {Bodenheimer}, \&
  {Laughlin}}]{Batygin2016}
{Batygin}, K., {Bodenheimer}, P.~H., \& {Laughlin}, G.~P. 2016, \apj, 829, 114

\bibitem[{{Batygin} \& {Stevenson}(2010)}]{Batygin2010}
{Batygin}, K. \& {Stevenson}, D.~J. 2010, \apjl, 714, L238

\bibitem[{{Boss}(1997)}]{Boss1997}
{Boss}, A.~P. 1997, Science, 276, 1836

\bibitem[{Boss {et~al.}(2005)Boss, Butler, Hubbard, Ianna, Kürster, Lissauer,
  Mayor, Meech, Mignard, Penny, \& et~al.}]{boss2005}
Boss, A.~P., Butler, R.~P., Hubbard, W.~B., {et~al.} 2005, Proceedings of the
  International Astronomical Union, 1, 183–186

\bibitem[{{Brahm} {et~al.}(2017){Brahm}, {Jord{\'a}n}, {Hartman}, \&
  {Bakos}}]{zaspe}
{Brahm}, R., {Jord{\'a}n}, A., {Hartman}, J., \& {Bakos}, G. 2017, \mnras, 467,
  971

\bibitem[{{Buchhave} {et~al.}(2010){Buchhave}, {Bakos}, {Hartman}, {Torres},
  {Kov{\'a}cs}, {Latham}, {Noyes}, {Esquerdo}, {Everett}, {Howard}, {Marcy},
  {Fischer}, {Johnson}, {Andersen}, {F{\H{u}}r{\'e}sz}, {Perumpilly},
  {Sasselov}, {Stefanik}, {B{\'e}ky}, {L{\'a}z{\'a}r}, {Papp}, \&
  {S{\'a}ri}}]{Buchhave2010}
{Buchhave}, L.~A., {Bakos}, G.~{\'A}., {Hartman}, J.~D., {et~al.} 2010, \apj,
  720, 1118

\bibitem[{{Buchhave} {et~al.}(2014){Buchhave}, {Bizzarro}, {Latham},
  {Sasselov}, {Cochran}, {Endl}, {Isaacson}, {Juncher}, \&
  {Marcy}}]{Buchhave2014}
{Buchhave}, L.~A., {Bizzarro}, M., {Latham}, D.~W., {et~al.} 2014, \nat, 509,
  593

\bibitem[{Buchhave {et~al.}(2012)Buchhave, Latham, Johansen, Bizzarro, Torres,
  Rowe, Batalha, Borucki, Brugamyer, Caldwell, Bryson, Ciardi, Cochran, Endl,
  Esquerdo, Ford, Geary, Gilliland, Hansen, Isaacson, Laird, Lucas, Marcy,
  Morse, Robertson, Shporer, Stefanik, Still, \& Quinn}]{Buchhave2012}
Buchhave, L.~A., Latham, D., Johansen, A., {et~al.} 2012, Nature, 486, 375

\bibitem[{{Cannon} \& {Pickering}(1993)}]{hd_id}
{Cannon}, A.~J. \& {Pickering}, E.~C. 1993, VizieR Online Data Catalog,
  III/135A

\bibitem[{{Castelli} \& {Kurucz}(2003)}]{zaspegrid}
{Castelli}, F. \& {Kurucz}, R.~L. 2003, in Modelling of Stellar Atmospheres,
  ed. N.~{Piskunov}, W.~W. {Weiss}, \& D.~F. {Gray}, Vol. 210, A20

\bibitem[{{Chabrier} {et~al.}(2014){Chabrier}, {Johansen}, {Janson}, \&
  {Rafikov}}]{Chabrier2014}
{Chabrier}, G., {Johansen}, A., {Janson}, M., \& {Rafikov}, R. 2014, in
  Protostars and Planets VI, ed. H.~{Beuther}, R.~S. {Klessen}, C.~P.
  {Dullemond}, \& T.~{Henning}, 619

\bibitem[{Chakraborty {et~al.}(2014)Chakraborty, Mahadevan, Roy, Dixit,
  Richardson, Dongre, Pathan, Chaturvedi, Shah, Ubale, \&
  Anandarao}]{Chakraborty_2014}
Chakraborty, A., Mahadevan, S., Roy, A., {et~al.} 2014, Publications of the
  Astronomical Society of the Pacific, 126, 133

\bibitem[{{Chakraborty} {et~al.}(2018){Chakraborty}, {Thapa}, {Kumar},
  {Neelam}, {Sharma}, \& {Roy}}]{paras2}
{Chakraborty}, A., {Thapa}, N., {Kumar}, K., {et~al.} 2018, in Society of
  Photo-Optical Instrumentation Engineers (SPIE) Conference Series, Vol. 10702,
  Ground-based and Airborne Instrumentation for Astronomy VII, ed. C.~J.
  {Evans}, L.~{Simard}, \& H.~{Takami}, 107026G

\bibitem[{Chaturvedi {et~al.}(2016)Chaturvedi, Chakraborty, Anandarao, Roy, \&
  Mahadevan}]{10.1093/mnras/stw1560}
Chaturvedi, P., Chakraborty, A., Anandarao, B.~G., Roy, A., \& Mahadevan, S.
  2016, Monthly Notices of the Royal Astronomical Society, 462, 554

\bibitem[{Chaturvedi {et~al.}(2018)Chaturvedi, Sharma, Chakraborty, Anandarao,
  \& Prasad}]{Chaturvedi_2018}
Chaturvedi, P., Sharma, R., Chakraborty, A., Anandarao, B.~G., \& Prasad, N. J.
  S. S.~V. 2018, The Astronomical Journal, 156, 27

\bibitem[{{Choi} {et~al.}(2016){Choi}, {Dotter}, {Conroy}, {Cantiello},
  {Paxton}, \& {Johnson}}]{mist_choi}
{Choi}, J., {Dotter}, A., {Conroy}, C., {et~al.} 2016, \apj, 823, 102

\bibitem[{{Ciardi} {et~al.}(2015){Ciardi}, {Beichman}, {Horch}, \&
  {Howell}}]{ciardi2015}
{Ciardi}, D.~R., {Beichman}, C.~A., {Horch}, E.~P., \& {Howell}, S.~B. 2015,
  \apj, 805, 16

\bibitem[{{Claret}(2017)}]{Claret_tess}
{Claret}, A. 2017, \aap, 600, A30

\bibitem[{{Claret} \& {Bloemen}(2011)}]{Claret}
{Claret}, A. \& {Bloemen}, S. 2011, \aap, 529, A75

\bibitem[{{Cutri} {et~al.}(2003){Cutri}, {Skrutskie}, {van Dyk}, {Beichman},
  {Carpenter}, {Chester}, {Cambresy}, {Evans}, {Fowler}, {Gizis}, {Howard},
  {Huchra}, {Jarrett}, {Kopan}, {Kirkpatrick}, {Light}, {Marsh}, {McCallon},
  {Schneider}, {Stiening}, {Sykes}, {Weinberg}, {Wheaton}, {Wheelock}, \&
  {Zacarias}}]{JHK}
{Cutri}, R.~M., {Skrutskie}, M.~F., {van Dyk}, S., {et~al.} 2003, VizieR Online
  Data Catalog, II/246

\bibitem[{{Cutri} {et~al.}(2021){Cutri}, {Wright}, {Conrow}, {Fowler},
  {Eisenhardt}, {Grillmair}, {Kirkpatrick}, {Masci}, {McCallon}, {Wheelock},
  {Fajardo-Acosta}, {Yan}, {Benford}, {Harbut}, {Jarrett}, {Lake}, {Leisawitz},
  {Ressler}, {Stanford}, {Tsai}, {Liu}, {Helou}, {Mainzer}, {Gettngs},
  {Gonzalez}, {Hoffman}, {Marsh}, {Padgett}, {Skrutskie}, {Beck}, {Papin}, \&
  {Wittman}}]{ALLWISE}
{Cutri}, R.~M., {Wright}, E.~L., {Conrow}, T., {et~al.} 2021, VizieR Online
  Data Catalog, II/328

\bibitem[{{Dawson} \& {Johnson}(2018)}]{dawson2018}
{Dawson}, R.~I. \& {Johnson}, J.~A. 2018, \araa, 56, 175

\bibitem[{{Dawson} \& {Murray-Clay}(2013)}]{Dawson-Murray-Clay-2013}
{Dawson}, R.~I. \& {Murray-Clay}, R.~A. 2013, \apjl, 767, L24

\bibitem[{{Dekany} {et~al.}(2013){Dekany}, {Roberts}, {Burruss}, {Bouchez},
  {Truong}, {Baranec}, {Guiwits}, {Hale}, {Angione}, {Trinh}, {Zolkower},
  {Shelton}, {Palmer}, {Henning}, {Croner}, {Troy}, {McKenna}, {Tesch},
  {Hildebrandt}, \& {Milburn}}]{dekany2013}
{Dekany}, R., {Roberts}, J., {Burruss}, R., {et~al.} 2013, \apj, 776, 130

\bibitem[{{Dong} {et~al.}(2021){Dong}, {Huang}, {Zhou}, {Dawson}, {Rodriguez},
  {Eastman}, {Collins}, {Quinn}, {Shporer}, {Triaud}, {Wang}, {Beatty},
  {Jackson}, {Collins}, {Abe}, {Suarez}, {Crouzet}, {M{\'e}karnia},
  {Dransfield}, {Jensen}, {Stockdale}, {Barkaoui}, {Heitzmann}, {Wright},
  {Addison}, {Wittenmyer}, {Okumura}, {Bowler}, {Horner}, {Kane}, {Kielkopf},
  {Liu}, {Plavchan}, {Mengel}, {Ricker}, {Vanderspek}, {Latham}, {Seager},
  {Winn}, {Jenkins}, {Christiansen}, \& {Paegert}}]{toi3362b}
{Dong}, J., {Huang}, C.~X., {Zhou}, G., {et~al.} 2021, \apjl, 920, L16

\bibitem[{{Dotter}(2016)}]{mist_dotter}
{Dotter}, A. 2016, \apjs, 222, 8

\bibitem[{{Eastman} {et~al.}(2019){Eastman}, {Rodriguez}, {Agol}, {Stassun},
  {Beatty}, {Vanderburg}, {Gaudi}, {Collins}, \& {Luger}}]{exofast}
{Eastman}, J.~D., {Rodriguez}, J.~E., {Agol}, E., {et~al.} 2019, arXiv
  e-prints, arXiv:1907.09480

\bibitem[{F\H{u}r\'esz(2008)}]{gaborthesis}
F\H{u}r\'esz, G. 2008, PhD thesis, University of Szeged, Hungary

\bibitem[{{Ford}(2006)}]{Gelman_rubin2006}
{Ford}, E.~B. 2006, \apj, 642, 505

\bibitem[{{Furlan} {et~al.}(2017){Furlan}, {Ciardi}, {Everett}, {Saylors},
  {Teske}, {Horch}, {Howell}, {van Belle}, {Hirsch}, {Gautier}, {Adams},
  {Barrado}, {Cartier}, {Dressing}, {Dupree}, {Gilliland}, {Lillo-Box},
  {Lucas}, \& {Wang}}]{furlan2017}
{Furlan}, E., {Ciardi}, D.~R., {Everett}, M.~E., {et~al.} 2017, \aj, 153, 71

\bibitem[{{Gaia Collaboration} {et~al.}(2021){Gaia Collaboration}, {Brown},
  {Vallenari}, {Prusti}, {de Bruijne}, {Babusiaux}, {Biermann}, {Creevey},
  {Evans}, {Eyer}, {Hutton}, {Jansen}, {Jordi}, {Klioner}, {Lammers},
  {Lindegren}, {Luri}, {Mignard}, {Panem}, {Pourbaix}, {Randich}, {Sartoretti},
  {Soubiran}, {Walton}, {Arenou}, {Bailer-Jones}, {Bastian}, {Cropper},
  {Drimmel}, {Katz}, {Lattanzi}, {van Leeuwen}, {Bakker}, {Cacciari},
  {Casta{\~n}eda}, {De Angeli}, {Ducourant}, {Fabricius}, {Fouesneau},
  {Fr{\'e}mat}, {Guerra}, {Guerrier}, {Guiraud}, {Jean-Antoine Piccolo},
  {Masana}, {Messineo}, {Mowlavi}, {Nicolas}, {Nienartowicz}, {Pailler},
  {Panuzzo}, {Riclet}, {Roux}, {Seabroke}, {Sordo}, {Tanga}, {Th{\'e}venin},
  {Gracia-Abril}, {Portell}, {Teyssier}, {Altmann}, {Andrae}, {Bellas-Velidis},
  {Benson}, {Berthier}, {Blomme}, {Brugaletta}, {Burgess}, {Busso}, {Carry},
  {Cellino}, {Cheek}, {Clementini}, {Damerdji}, {Davidson}, {Delchambre},
  {Dell'Oro}, {Fern{\'a}ndez-Hern{\'a}ndez}, {Galluccio}, {Garc{\'\i}a-Lario},
  {Garcia-Reinaldos}, {Gonz{\'a}lez-N{\'u}{\~n}ez}, {Gosset}, {Haigron},
  {Halbwachs}, {Hambly}, {Harrison}, {Hatzidimitriou}, {Heiter},
  {Hern{\'a}ndez}, {Hestroffer}, {Hodgkin}, {Holl}, {Jan{\ss}en}, {Jevardat de
  Fombelle}, {Jordan}, {Krone-Martins}, {Lanzafame}, {L{\"o}ffler}, {Lorca},
  {Manteiga}, {Marchal}, {Marrese}, {Moitinho}, {Mora}, {Muinonen}, {Osborne},
  {Pancino}, {Pauwels}, {Petit}, {Recio-Blanco}, {Richards}, {Riello},
  {Rimoldini}, {Robin}, {Roegiers}, {Rybizki}, {Sarro}, {Siopis}, {Smith},
  {Sozzetti}, {Ulla}, {Utrilla}, {van Leeuwen}, {van Reeven}, {Abbas}, {Abreu
  Aramburu}, {Accart}, {Aerts}, {Aguado}, {Ajaj}, {Altavilla}, {{\'A}lvarez},
  {{\'A}lvarez Cid-Fuentes}, {Alves}, {Anderson}, {Anglada Varela}, {Antoja},
  {Audard}, {Baines}, {Baker}, {Balaguer-N{\'u}{\~n}ez}, {Balbinot}, {Balog},
  {Barache}, {Barbato}, {Barros}, {Barstow}, {Bartolom{\'e}}, {Bassilana},
  {Bauchet}, {Baudesson-Stella}, {Becciani}, {Bellazzini}, {Bernet}, {Bertone},
  {Bianchi}, {Blanco-Cuaresma}, {Boch}, {Bombrun}, {Bossini}, {Bouquillon},
  {Bragaglia}, {Bramante}, {Breedt}, {Bressan}, {Brouillet}, {Bucciarelli},
  {Burlacu}, {Busonero}, {Butkevich}, {Buzzi}, {Caffau}, {Cancelliere},
  {C{\'a}novas}, {Cantat-Gaudin}, {Carballo}, {Carlucci}, {Carnerero},
  {Carrasco}, {Casamiquela}, {Castellani}, {Castro-Ginard}, {Castro Sampol},
  {Chaoul}, {Charlot}, {Chemin}, {Chiavassa}, {Cioni}, {Comoretto}, {Cooper},
  {Cornez}, {Cowell}, {Crifo}, {Crosta}, {Crowley}, {Dafonte}, {Dapergolas},
  {David}, {David}, {de Laverny}, {De Luise}, {De March}, {De Ridder}, {de
  Souza}, {de Teodoro}, {de Torres}, {del Peloso}, {del Pozo}, {Delbo},
  {Delgado}, {Delgado}, {Delisle}, {Di Matteo}, {Diakite}, {Diener},
  {Distefano}, {Dolding}, {Eappachen}, {Edvardsson}, {Enke}, {Esquej}, {Fabre},
  {Fabrizio}, {Faigler}, {Fedorets}, {Fernique}, {Fienga}, {Figueras},
  {Fouron}, {Fragkoudi}, {Fraile}, {Franke}, {Gai}, {Garabato},
  {Garcia-Gutierrez}, {Garc{\'\i}a-Torres}, {Garofalo}, {Gavras}, {Gerlach},
  {Geyer}, {Giacobbe}, {Gilmore}, {Girona}, {Giuffrida}, {Gomel}, {Gomez},
  {Gonzalez-Santamaria}, {Gonz{\'a}lez-Vidal}, {Granvik},
  {Guti{\'e}rrez-S{\'a}nchez}, {Guy}, {Hauser}, {Haywood}, {Helmi}, {Hidalgo},
  {Hilger}, {H{\l}adczuk}, {Hobbs}, {Holland}, {Huckle}, {Jasniewicz},
  {Jonker}, {Juaristi Campillo}, {Julbe}, {Karbevska}, {Kervella}, {Khanna},
  {Kochoska}, {Kontizas}, {Kordopatis}, {Korn}, {Kostrzewa-Rutkowska},
  {Kruszy{\'n}ska}, {Lambert}, {Lanza}, {Lasne}, {Le Campion}, {Le Fustec},
  {Lebreton}, {Lebzelter}, {Leccia}, {Leclerc}, {Lecoeur-Taibi}, {Liao},
  {Licata}, {Lindstr{\o}m}, {Lister}, {Livanou}, {Lobel}, {Madrero Pardo},
  {Managau}, {Mann}, {Marchant}, {Marconi}, {Marcos Santos}, {Marinoni},
  {Marocco}, {Marshall}, {Martin Polo}, {Mart{\'\i}n-Fleitas}, {Masip},
  {Massari}, {Mastrobuono-Battisti}, {Mazeh}, {McMillan}, {Messina},
  {Michalik}, {Millar}, {Mints}, {Molina}, {Molinaro}, {Moln{\'a}r},
  {Montegriffo}, {Mor}, {Morbidelli}, {Morel}, {Morris}, {Mulone}, {Munoz},
  {Muraveva}, {Murphy}, {Musella}, {Noval}, {Ord{\'e}novic}, {Orr{\`u}},
  {Osinde}, {Pagani}, {Pagano}, {Palaversa}, {Palicio}, {Panahi}, {Pawlak},
  {Pe{\~n}alosa Esteller}, {Penttil{\"a}}, {Piersimoni}, {Pineau}, {Plachy},
  {Plum}, {Poggio}, {Poretti}, {Poujoulet}, {Pr{\v{s}}a}, {Pulone}, {Racero},
  {Ragaini}, {Rainer}, {Raiteri}, {Rambaux}, {Ramos}, {Ramos-Lerate}, {Re
  Fiorentin}, {Regibo}, {Reyl{\'e}}, {Ripepi}, {Riva}, {Rixon}, {Robichon},
  {Robin}, {Roelens}, {Rohrbasser}, {Romero-G{\'o}mez}, {Rowell}, {Royer},
  {Rybicki}, {Sadowski}, {Sagrist{\`a} Sell{\'e}s}, {Sahlmann}, {Salgado},
  {Salguero}, {Samaras}, {Sanchez Gimenez}, {Sanna}, {Santove{\~n}a},
  {Sarasso}, {Schultheis}, {Sciacca}, {Segol}, {Segovia}, {S{\'e}gransan},
  {Semeux}, {Shahaf}, {Siddiqui}, {Siebert}, {Siltala}, {Slezak}, {Smart},
  {Solano}, {Solitro}, {Souami}, {Souchay}, {Spagna}, {Spoto}, {Steele},
  {Steidelm{\"u}ller}, {Stephenson}, {S{\"u}veges}, {Szabados}, {Szegedi-Elek},
  {Taris}, {Tauran}, {Taylor}, {Teixeira}, {Thuillot}, {Tonello}, {Torra},
  {Torra}, {Turon}, {Unger}, {Vaillant}, {van Dillen}, {Vanel}, {Vecchiato},
  {Viala}, {Vicente}, {Voutsinas}, {Weiler}, {Wevers}, {Wyrzykowski}, {Yoldas},
  {Yvard}, {Zhao}, {Zorec}, {Zucker}, {Zurbach}, \& {Zwitter}}]{gaiaedr3}
{Gaia Collaboration}, {Brown}, A.~G.~A., {Vallenari}, A., {et~al.} 2021, \aap,
  649, A1

\bibitem[{{Gaia Collaboration} {et~al.}(2022){Gaia Collaboration}, {Vallenari},
  {Brown}, {Prusti}, {de Bruijne}, {Arenou}, {Babusiaux}, {Biermann},
  {Creevey}, {Ducourant}, {Evans}, {Eyer}, {Guerra}, {Hutton}, {Jordi},
  {Klioner}, {Lammers}, {Lindegren}, {Luri}, {Mignard}, {Panem}, {Pourbaix},
  {Randich}, {Sartoretti}, {Soubiran}, {Tanga}, {Walton}, {Bailer-Jones},
  {Bastian}, {Drimmel}, {Jansen}, {Katz}, {Lattanzi}, {van Leeuwen}, {Bakker},
  {Cacciari}, {Casta{\~n}eda}, {De Angeli}, {Fabricius}, {Fouesneau},
  {Fr{\'e}mat}, {Galluccio}, {Guerrier}, {Heiter}, {Masana}, {Messineo},
  {Mowlavi}, {Nicolas}, {Nienartowicz}, {Pailler}, {Panuzzo}, {Riclet}, {Roux},
  {Seabroke}, {Sordo{\o}rcit}, {Th{\'e}venin}, {Gracia-Abril}, {Portell},
  {Teyssier}, {Altmann}, {Andrae}, {Audard}, {Bellas-Velidis}, {Benson},
  {Berthier}, {Blomme}, {Burgess}, {Busonero}, {Busso}, {C{\'a}novas}, {Carry},
  {Cellino}, {Cheek}, {Clementini}, {Damerdji}, {Davidson}, {de Teodoro},
  {Nu{\~n}ez Campos}, {Delchambre}, {Dell'Oro}, {Esquej},
  {Fern{\'a}ndez-Hern{\'a}ndez}, {Fraile}, {Garabato}, {Garc{\'\i}a-Lario},
  {Gosset}, {Haigron}, {Halbwachs}, {Hambly}, {Harrison}, {Hern{\'a}ndez},
  {Hestroffer}, {Hodgkin}, {Holl}, {Jan{\ss}en}, {Jevardat de Fombelle},
  {Jordan}, {Krone-Martins}, {Lanzafame}, {L{\"o}ffler}, {Marchal}, {Marrese},
  {Moitinho}, {Muinonen}, {Osborne}, {Pancino}, {Pauwels}, {Recio-Blanco},
  {Reyl{\'e}}, {Riello}, {Rimoldini}, {Roegiers}, {Rybizki}, {Sarro}, {Siopis},
  {Smith}, {Sozzetti}, {Utrilla}, {van Leeuwen}, {Abbas}, {{\'A}brah{\'a}m},
  {Abreu Aramburu}, {Aerts}, {Aguado}, {Ajaj}, {Aldea-Montero}, {Altavilla},
  {{\'A}lvarez}, {Alves}, {Anders}, {Anderson}, {Anglada Varela}, {Antoja},
  {Baines}, {Baker}, {Balaguer-N{\'u}{\~n}ez}, {Balbinot}, {Balog}, {Barache},
  {Barbato}, {Barros}, {Barstow}, {Bartolom{\'e}}, {Bassilana}, {Bauchet},
  {Becciani}, {Bellazzini}, {Berihuete}, {Bernet}, {Bertone}, {Bianchi},
  {Binnenfeld}, {Blanco-Cuaresma}, {Blazere}, {Boch}, {Bombrun}, {Bossini},
  {Bouquillon}, {Bragaglia}, {Bramante}, {Breedt}, {Bressan}, {Brouillet},
  {Brugaletta}, {Bucciarelli}, {Burlacu}, {Butkevich}, {Buzzi}, {Caffau},
  {Cancelliere}, {Cantat-Gaudin}, {Carballo}, {Carlucci}, {Carnerero},
  {Carrasco}, {Casamiquela}, {Castellani}, {Castro-Ginard}, {Chaoul},
  {Charlot}, {Chemin}, {Chiaramida}, {Chiavassa}, {Chornay}, {Comoretto},
  {Contursi}, {Cooper}, {Cornez}, {Cowell}, {Crifo}, {Cropper}, {Crosta},
  {Crowley}, {Dafonte}, {Dapergolas}, {David}, {David}, {de Laverny}, {De
  Luise}, {De March}, {De Ridder}, {de Souza}, {de Torres}, {del Peloso}, {del
  Pozo}, {Delbo}, {Delgado}, {Delisle}, {Demouchy}, {Dharmawardena}, {Di
  Matteo}, {Diakite}, {Diener}, {Distefano}, {Dolding}, {Edvardsson}, {Enke},
  {Fabre}, {Fabrizio}, {Faigler}, {Fedorets}, {Fernique}, {Fienga}, {Figueras},
  {Fournier}, {Fouron}, {Fragkoudi}, {Gai}, {Garcia-Gutierrez},
  {Garcia-Reinaldos}, {Garc{\'\i}a-Torres}, {Garofalo}, {Gavel}, {Gavras},
  {Gerlach}, {Geyer}, {Giacobbe}, {Gilmore}, {Girona}, {Giuffrida}, {Gomel},
  {Gomez}, {Gonz{\'a}lez-N{\'u}{\~n}ez}, {Gonz{\'a}lez-Santamar{\'\i}a},
  {Gonz{\'a}lez-Vidal}, {Granvik}, {Guillout}, {Guiraud},
  {Guti{\'e}rrez-S{\'a}nchez}, {Guy}, {Hatzidimitriou}, {Hauser}, {Haywood},
  {Helmer}, {Helmi}, {Sarmiento}, {Hidalgo}, {Hilger}, {H{\l}adczuk}, {Hobbs},
  {Holland}, {Huckle}, {Jardine}, {Jasniewicz}, {Jean-Antoine Piccolo},
  {Jim{\'e}nez-Arranz}, {Jorissen}, {Juaristi Campillo}, {Julbe}, {Karbevska},
  {Kervella}, {Khanna}, {Kontizas}, {Kordopatis}, {Korn}, {K{\'o}sp{\'a}l},
  {Kostrzewa-Rutkowska}, {Kruszy{\'n}ska}, {Kun}, {Laizeau}, {Lambert},
  {Lanza}, {Lasne}, {Le Campion}, {Lebreton}, {Lebzelter}, {Leccia}, {Leclerc},
  {Lecoeur-Taibi}, {Liao}, {Licata}, {Lindstr{\o}m}, {Lister}, {Livanou},
  {Lobel}, {Lorca}, {Loup}, {Madrero Pardo}, {Magdaleno Romeo}, {Managau},
  {Mann}, {Manteiga}, {Marchant}, {Marconi}, {Marcos}, {Marcos Santos},
  {Mar{\'\i}n Pina}, {Marinoni}, {Marocco}, {Marshall}, {Polo},
  {Mart{\'\i}n-Fleitas}, {Marton}, {Mary}, {Masip}, {Massari},
  {Mastrobuono-Battisti}, {Mazeh}, {McMillan}, {Messina}, {Michalik}, {Millar},
  {Mints}, {Molina}, {Molinaro}, {Moln{\'a}r}, {Monari}, {Mongui{\'o}},
  {Montegriffo}, {Montero}, {Mor}, {Mora}, {Morbidelli}, {Morel}, {Morris},
  {Muraveva}, {Murphy}, {Musella}, {Nagy}, {Noval}, {Oca{\~n}a}, {Ogden},
  {Ordenovic}, {Osinde}, {Pagani}, {Pagano}, {Palaversa}, {Palicio},
  {Pallas-Quintela}, {Panahi}, {Payne-Wardenaar}, {Pe{\~n}alosa Esteller},
  {Penttil{\"a}}, {Pichon}, {Piersimoni}, {Pineau}, {Plachy}, {Plum}, {Poggio},
  {Pr{\v{s}}a}, {Pulone}, {Racero}, {Ragaini}, {Rainer}, {Raiteri}, {Rambaux},
  {Ramos}, {Ramos-Lerate}, {Re Fiorentin}, {Regibo}, {Richards}, {Rios Diaz},
  {Ripepi}, {Riva}, {Rix}, {Rixon}, {Robichon}, {Robin}, {Robin}, {Roelens},
  {Rogues}, {Rohrbasser}, {Romero-G{\'o}mez}, {Rowell}, {Royer}, {Ruz Mieres},
  {Rybicki}, {Sadowski}, {S{\'a}ez N{\'u}{\~n}ez}, {Sagrist{\`a} Sell{\'e}s},
  {Sahlmann}, {Salguero}, {Samaras}, {Sanchez Gimenez}, {Sanna},
  {Santove{\~n}a}, {Sarasso}, {Schultheis}, {Sciacca}, {Segol}, {Segovia},
  {S{\'e}gransan}, {Semeux}, {Shahaf}, {Siddiqui}, {Siebert}, {Siltala},
  {Silvelo}, {Slezak}, {Slezak}, {Smart}, {Snaith}, {Solano}, {Solitro},
  {Souami}, {Souchay}, {Spagna}, {Spina}, {Spoto}, {Steele},
  {Steidelm{\"u}ller}, {Stephenson}, {S{\"u}veges}, {Surdej}, {Szabados},
  {Szegedi-Elek}, {Taris}, {Taylo}, {Teixeira}, {Tolomei}, {Tonello}, {Torra},
  {Torra}, {Torralba Elipe}, {Trabucchi}, {Tsounis}, {Turon}, {Ulla}, {Unger},
  {Vaillant}, {van Dillen}, {van Reeven}, {Vanel}, {Vecchiato}, {Viala},
  {Vicente}, {Voutsinas}, {Weiler}, {Wevers}, {Wyrzykowski}, {Yoldas}, {Yvard},
  {Zhao}, {Zorec}, {Zucker}, \& {Zwitter}}]{gaiadr3}
{Gaia Collaboration}, {Vallenari}, A., {Brown}, A.~G.~A., {et~al.} 2022, arXiv
  e-prints, arXiv:2208.00211

\bibitem[{{Gelman} \& {Rubin}(1992)}]{Gelman_rubin1992}
{Gelman}, A. \& {Rubin}, D.~B. 1992, Statistical Science, 7, 457

\bibitem[{{Guenther} {et~al.}(2009){Guenther}, {Hartmann}, {Esposito},
  {Hatzes}, {Cusano}, \& {Gandolfi}}]{2009A&A...507.1659G}
{Guenther}, E.~W., {Hartmann}, M., {Esposito}, M., {et~al.} 2009, \aap, 507,
  1659

\bibitem[{{Hatzes} \& {Rauer}(2015)}]{artie2015}
{Hatzes}, A.~P. \& {Rauer}, H. 2015, \apjl, 810, L25

\bibitem[{{Hayward} {et~al.}(2001){Hayward}, {Brandl}, {Pirger}, {Blacken},
  {Gull}, {Schoenwald}, \& {Houck}}]{hayward2001}
{Hayward}, T.~L., {Brandl}, B., {Pirger}, B., {et~al.} 2001, \pasp, 113, 105

\bibitem[{{Henden} {et~al.}(2016){Henden}, {Templeton}, {Terrell}, {Smith},
  {Levine}, \& {Welch}}]{APASS}
{Henden}, A.~A., {Templeton}, M., {Terrell}, D., {et~al.} 2016, VizieR Online
  Data Catalog, II/336

\bibitem[{{H{\o}g} {et~al.}(2000){H{\o}g}, {Fabricius}, {Makarov}, {Urban},
  {Corbin}, {Wycoff}, {Bastian}, {Schwekendiek}, \& {Wicenec}}]{tycho}
{H{\o}g}, E., {Fabricius}, C., {Makarov}, V.~V., {et~al.} 2000, \aap, 355, L27

\bibitem[{Jenkins {et~al.}(2016)Jenkins, Twicken, McCauliff, Campbell,
  Sanderfer, Lung, Mansouri-Samani, Girouard, Tenenbaum, Klaus, Smith,
  Caldwell, Chacon, Henze, Heiges, Latham, Morgan, Swade, Rinehart, \&
  Vanderspek}]{spoc}
Jenkins, J., Twicken, J., McCauliff, S., {et~al.} 2016, in Software and Cyber
  infrastructure for Astronomy IV

\bibitem[{{Johns-Krull} {et~al.}(2008){Johns-Krull}, {McCullough}, {Burke},
  {Valenti}, {Janes}, {Heasley}, {Prato}, {Bissinger}, {Fleenor}, {Foote},
  {Garcia-Melendo}, {Gary}, {Howell}, {Mallia}, {Masi}, \& {Vanmunster}}]{xo3}
{Johns-Krull}, C.~M., {McCullough}, P.~R., {Burke}, C.~J., {et~al.} 2008, \apj,
  677, 657

\bibitem[{{Kervella} {et~al.}(2019){Kervella}, {Arenou}, {Mignard}, \&
  {Th{\'e}venin}}]{gaia_proper_motion}
{Kervella}, P., {Arenou}, F., {Mignard}, F., \& {Th{\'e}venin}, F. 2019, \aap,
  623, A72

\bibitem[{{Khandelwal} {et~al.}(2022){Khandelwal}, {Chaturvedi}, {Chakraborty},
  {Sharma}, {Guenther}, {Persson}, {Fridlund}, {Hatzes}, {Prasad}, {Esposito},
  {Chamarthi}, {Nayak}, {Dishendra}, \& {Howell}}]{toi1789}
{Khandelwal}, A., {Chaturvedi}, P., {Chakraborty}, A., {et~al.} 2022, \mnras,
  509, 3339

\bibitem[{{Komacek} \& {Youdin}(2017)}]{Komacek2017}
{Komacek}, T.~D. \& {Youdin}, A.~N. 2017, \apj, 844, 94

\bibitem[{{Kurucz}(1979)}]{Kurucz}
{Kurucz}, R.~L. 1979, \apjs, 40, 1

\bibitem[{{Kurucz}(1992)}]{kurucz1992}
{Kurucz}, R.~L. 1992, in The Stellar Populations of Galaxies, ed. B.~{Barbuy}
  \& A.~{Renzini}, Vol. 149, 225

\bibitem[{{Lecavelier des Etangs} \& {Lissauer}(2022)}]{IAU2022}
{Lecavelier des Etangs}, A. \& {Lissauer}, J.~J. 2022, \nar, 94, 101641

\bibitem[{{Lightkurve Collaboration} {et~al.}(2018){Lightkurve Collaboration},
  {Cardoso}, {Hedges}, {Gully-Santiago}, {Saunders}, {Cody}, {Barclay}, {Hall},
  {Sagear}, {Turtelboom}, {Zhang}, {Tzanidakis}, {Mighell}, {Coughlin}, {Bell},
  {Berta-Thompson}, {Williams}, {Dotson}, \& {Barentsen}}]{lightkurve}
{Lightkurve Collaboration}, {Cardoso}, J.~V.~d.~M., {Hedges}, C., {et~al.}
  2018, {Lightkurve: Kepler and TESS time series analysis in Python},
  Astrophysics Source Code Library

\bibitem[{{Lund} {et~al.}(2016){Lund}, {Chaplin}, {Casagrande}, {Silva
  Aguirre}, {Basu}, {Bieryla}, {Christensen-Dalsgaard}, {Latham}, {White},
  {Davies}, {Huber}, {Buchhave}, \& {Handberg}}]{Lund2016}
{Lund}, M.~N., {Chaplin}, W.~J., {Casagrande}, L., {et~al.} 2016, \pasp, 128,
  124204

\bibitem[{{Mandel} \& {Agol}(2002)}]{Mandel2002}
{Mandel}, K. \& {Agol}, E. 2002, \apjl, 580, L171

\bibitem[{{McLaughlin}(1924)}]{1924ApJ....60...22M}
{McLaughlin}, D.~B. 1924, \apj, 60, 22

\bibitem[{{Mordasini}(2020)}]{Mordasini2020}
{Mordasini}, C. 2020, \aap, 638, A52

\bibitem[{{Mordasini} {et~al.}(2012){Mordasini}, {Alibert}, {Klahr}, \&
  {Henning}}]{Mordasini2012}
{Mordasini}, C., {Alibert}, Y., {Klahr}, H., \& {Henning}, T. 2012, \aap, 547,
  A111

\bibitem[{{Ohta} {et~al.}(2005){Ohta}, {Taruya}, \& {Suto}}]{rm}
{Ohta}, Y., {Taruya}, A., \& {Suto}, Y. 2005, \apj, 622, 1118

\bibitem[{{Pollack} {et~al.}(1996){Pollack}, {Hubickyj}, {Bodenheimer},
  {Lissauer}, {Podolak}, \& {Greenzweig}}]{pollack1996}
{Pollack}, J.~B., {Hubickyj}, O., {Bodenheimer}, P., {et~al.} 1996, \icarus,
  124, 62

\bibitem[{{Pont} {et~al.}(2005){Pont}, {Bouchy}, {Melo}, {Santos}, {Mayor},
  {Queloz}, \& {Udry}}]{Pont2005}
{Pont}, F., {Bouchy}, F., {Melo}, C., {et~al.} 2005, \aap, 438, 1123

\bibitem[{{Rossiter}(1924)}]{1924ApJ....60...15R}
{Rossiter}, R.~A. 1924, \apj, 60, 15

\bibitem[{{Santos} {et~al.}(2017){Santos}, {Adibekyan}, {Figueira},
  {Andreasen}, {Barros}, {Delgado-Mena}, {Demangeon}, {Faria}, {Oshagh},
  {Sousa}, {Viana}, \& {Ferreira}}]{Santos2017}
{Santos}, N.~C., {Adibekyan}, V., {Figueira}, P., {et~al.} 2017, \aap, 603, A30

\bibitem[{{Sarkis} {et~al.}(2021){Sarkis}, {Mordasini}, {Henning}, {Marleau},
  \& {Molli{\`e}re}}]{Sarkis2021}
{Sarkis}, P., {Mordasini}, C., {Henning}, T., {Marleau}, G.~D., \&
  {Molli{\`e}re}, P. 2021, \aap, 645, A79

\bibitem[{{Saumon} {et~al.}(1995){Saumon}, {Chabrier}, \& {van
  Horn}}]{Saumon1995}
{Saumon}, D., {Chabrier}, G., \& {van Horn}, H.~M. 1995, \apjs, 99, 713

\bibitem[{{Schlafly} \& {Finkbeiner}(2011)}]{extinction}
{Schlafly}, E.~F. \& {Finkbeiner}, D.~P. 2011, \apj, 737, 103

\bibitem[{{Schlaufman}(2018)}]{Schlaufman2018}
{Schlaufman}, K.~C. 2018, \apj, 853, 37

\bibitem[{{Schneider} {et~al.}(2011){Schneider}, {Dedieu}, {Le Sidaner},
  {Savalle}, \& {Zolotukhin}}]{Schneider2011}
{Schneider}, J., {Dedieu}, C., {Le Sidaner}, P., {Savalle}, R., \&
  {Zolotukhin}, I. 2011, \aap, 532, A79

\bibitem[{Sharma \& Chakraborty(2021)}]{uar}
Sharma, R. \& Chakraborty, A.~G. 2021, Journal of Astronomical Telescopes,
  Instruments, and Systems, 7, 1

\bibitem[{{Smith} {et~al.}(2012){Smith}, {Stumpe}, {Van Cleve}, {Jenkins},
  {Barclay}, {Fanelli}, {Girouard}, {Kolodziejczak}, {McCauliff}, {Morris}, \&
  {Twicken}}]{smith_2012}
{Smith}, J.~C., {Stumpe}, M.~C., {Van Cleve}, J.~E., {et~al.} 2012, \pasp, 124,
  1000

\bibitem[{Southworth(2011)}]{TEPcat}
Southworth, J. 2011, Monthly Notices of the Royal Astronomical Society, 417,
  2166

\bibitem[{{Spiegel} {et~al.}(2011){Spiegel}, {Burrows}, \&
  {Milsom}}]{Spiegel2011}
{Spiegel}, D.~S., {Burrows}, A., \& {Milsom}, J.~A. 2011, \apj, 727, 57

\bibitem[{{Stassun} {et~al.}(2018){Stassun}, {Oelkers}, {Pepper}, {Paegert},
  {De Lee}, {Torres}, {Latham}, {Charpinet}, {Dressing}, {Huber}, {Kane},
  {L{\'e}pine}, {Mann}, {Muirhead}, {Rojas-Ayala}, {Silvotti}, {Fleming},
  {Levine}, \& {Plavchan}}]{2018AJ....156..102S}
{Stassun}, K.~G., {Oelkers}, R.~J., {Pepper}, J., {et~al.} 2018, \aj, 156, 102

\bibitem[{{Stassun} \& {Torres}(2016)}]{SED1}
{Stassun}, K.~G. \& {Torres}, G. 2016, \apjl, 831, L6

\bibitem[{Stumpe {et~al.}(2014)Stumpe, Smith, Catanzarite, Cleve, Jenkins,
  Twicken, \& Girouard}]{Stumpe_2014}
Stumpe, M.~C., Smith, J.~C., Catanzarite, J.~H., {et~al.} 2014, Publications of
  the Astronomical Society of the Pacific, 126, 100

\bibitem[{Thompson(1990)}]{Thompson1990}
Thompson, S.~L. 1990

\bibitem[{{Thorngren} \& {Fortney}(2018)}]{Thorngren2018}
{Thorngren}, D.~P. \& {Fortney}, J.~J. 2018, \aj, 155, 214

\bibitem[{{Thorngren} {et~al.}(2016){Thorngren}, {Fortney}, {Murray-Clay}, \&
  {Lopez}}]{Thorngren2016}
{Thorngren}, D.~P., {Fortney}, J.~J., {Murray-Clay}, R.~A., \& {Lopez}, E.~D.
  2016, \apj, 831, 64

\bibitem[{{Udry}(2010)}]{udry2010}
{Udry}, S. 2010, in In the Spirit of Lyot 2010, ed. A.~{Boccaletti}, E11

\bibitem[{{Ulmer-Moll} {et~al.}(2022){Ulmer-Moll}, {Lendl}, {Gill},
  {Villanueva}, {Hobson}, {Bouchy}, {Brahm}, {Dragomir}, {Grieves},
  {Mordasini}, {Anderson}, {Acton}, {Bayliss}, {Bieryla}, {Burleigh},
  {Casewell}, {Chaverot}, {Eigm{\"u}ller}, {Feliz}, {Gaudi}, {Gillen}, {Goad},
  {Gupta}, {G{\"u}nther}, {Henderson}, {Henning}, {Jenkins}, {Jones},
  {Jord{\'a}n}, {Kendall}, {Latham}, {Mireles}, {Moyano}, {Nadol}, {Osborn},
  {Pepper}, {Pinto}, {Psaridi}, {Queloz}, {Quinn}, {Rojas}, {Sarkis},
  {Schlecker}, {Tilbrook}, {Torres}, {Trifonov}, {Udry}, {Vines}, {West},
  {Wheatley}, {Yao}, {Zhao}, \& {Zhou}}]{Ulmer-Moll2022}
{Ulmer-Moll}, S., {Lendl}, M., {Gill}, S., {et~al.} 2022, \aap, 666, A46

\bibitem[{{Zechmeister} \& {K{\"u}rster}(2009)}]{periodogram}
{Zechmeister}, M. \& {K{\"u}rster}, M. 2009, \aap, 496, 577

\bibitem[{{Zhou} {et~al.}(2019){Zhou}, {Bakos}, {Bayliss}, {Bento}, {Bhatti},
  {Brahm}, {Csubry}, {Espinoza}, {Hartman}, {Henning}, {Jord{\'a}n}, {Mancini},
  {Penev}, {Rabus}, {Sarkis}, {Suc}, {de Val-Borro}, {Rodriguez}, {Osip},
  {Kedziora-Chudczer}, {Bailey}, {Tinney}, {Durkan}, {L{\'a}z{\'a}r}, {Papp},
  \& {S{\'a}ri}}]{hats70}
{Zhou}, G., {Bakos}, G.~{\'A}., {Bayliss}, D., {et~al.} 2019, \aj, 157, 31

\bibitem[{{Ziegler} {et~al.}(2020){Ziegler}, {Tokovinin}, {Brice{\~n}o},
  {Mang}, {Law}, \& {Mann}}]{ziegler2020}
{Ziegler}, C., {Tokovinin}, A., {Brice{\~n}o}, C., {et~al.} 2020, \aj, 159, 19

\end{thebibliography}

\begin{appendix}

\section{Tables}

\begin{table}[h!]
	\begin{center}

	\caption{\textcolor{black}{}Basic Stellar Parameters for TOI-4603}
	\label{tab:star_table}
	\begin{tabular}{cllc} 
		\hline
		\hline
		\noalign{\smallskip}
		Parameter & Description (unit) & Value & Source\\
		\noalign{\smallskip}
		\hline
		\noalign{\smallskip}
		$\alpha_{J2000}$ & Right Ascension & 05:35:27.82 & (1)\\
		$\delta_{J2000}$ & Declination & +21:17:39.62 & (1)\\
		$\mu_{\alpha}$ & PM in R.A. (mas yr$^{-1}$) & 0.102 $\pm$ 0.021 & (1)\\
		$\mu_{\delta}$ & PM in Dec (mas yr$^{-1}$) & -22.866 $\pm$ 0.011 & (1)\\
		$\pi$ & Parallax (mas) & 4.4613  $\pm$ 0.0195 & (1)\\
        $G$ & $Gaia$ G mag & 9.0831  $\pm$ 0.0027 & (1)\\
		$T$ & TESS T mag & 8.6554  $\pm$ 0.0062 & (2)\\
		$B_{T}$ & Tycho B mag  &9.964	 $\pm$ 0.026 & (3)\\
		$V_{T}$ & Tycho V mag & 9.273  $\pm$ 0.019  & (3)\\
		$B$   & APASS B-mag & 9.915  $\pm$ 0.03 & (6)\\
        $V$   & APASS V-mag  &9.421   $\pm$0.15 & (6)\\
        $g$   & SDSSg mag  & 9.968 $\pm$ 0.23 & (6)\\
        $r$   & SDSSr mag  &9.310 $\pm$ 0.18 & (6)\\
        $i$   & SDSSi mag  &8.976 $\pm$ 0.04 & (6)\\
		$J$   & 2MASS J mag & 8.089 $\pm$ 0.020 & (4)\\
		$H$   & 2MASS H mag & 7.788  $\pm$ 0.047 & (4)\\
		$K_{S}$ & 2MASS K$_S$ mag & 7.786 $\pm$ 0.017 & (4)\\
		$W1$  & WISE1 mag & 7.718  $\pm$ 0.028 & (5)\\
		$W2$  & WISE2 mag & 7.744   $\pm$ 0.02 & (5)\\
		$W3$  & WISE3 mag & 7.761  $\pm$ 0.02 & (5)\\
		$W4$  & WISE4 mag & 7.933  $\pm$ 0.198 & (5)\\
            $L_*$ & Luminosity (\(L_\odot\)) & 9.74 [9.65, 9.80] & (1)\\
            $T_{\rm eff}$ & Effective Temperature (K) & 6189 [6185, 6193]& (1)\\
            $\log{g}$ &Surface gravity (cgs) & 3.805 [3.801, 3.818] & (1)\\
            $[{\rm M/H}]$ & Metallicity (dex) & -0.236 [-0.239, -0.232]&(1)\\
            $M_*$&Mass (\(M_\odot\)) & 1.752 $\pm$ 0.088&(1)\\
            $R_*$ & Radius (\(R_\odot\))&2.722 $\pm$ 0.136&(1)\\
            $Age$ & Age (Gyr)& 1.98 [1.73, 2.22]& (1)\\
	\noalign{\smallskip}
	\hline
		
	\end{tabular}
	\end{center}
	Other Identifiers:
	\begin{center}
	    HD 245134$^7$\\
	    TIC 437856897$^2$\\
	    TYC 1309-1102-1$^3$\\
	    2MASS J05352782+2117396$^4$\\
	    $Gaia EDR3$ 3402980516507429888$^1$\\

	\end{center}
	\smallskip
	\hrule
	\smallskip
         \textcolor{black}{\textbf{Note:} The metallicity of TOI-4603 reported by Gaia is different from our spectroscopic analysis~(see section \ref{sec:spec_param}).}\\
	\textbf{References.} (1) \cite{gaiaedr3}, (2) \cite{2018AJ....156..102S}, (3) \cite{tycho}, (4) \cite{JHK}, (5) \cite{ALLWISE}, (6) \cite{APASS}, (7) \cite{hd_id}\\
	
\end{table}
\newpage
\onecolumn
\begin{center}
\begin{longtable}[c]{ccccccc}
    \caption{\textcolor{black}{}Radial Velocity measurements of TOI-4603. }
	\label{tab:rv_table}\\ \hline
	    \noalign{\smallskip}
		
		BJD$_{TDB}$& Relative-RV & $\sigma$-RV & BIS & $\sigma$-BIS & EXP-TIME&Instrument \\
		\vspace{0.1cm}\\
		Days & m s$^{-1}$ & m s$^{-1}$ & m s$^{-1}$ & m s$^{-1}$&s&\\\hline
	    \noalign{\smallskip}
2459591.245018 & 1301.87   & 57.30      &  -2207.44  &   268.35& 1800 & PARAS \\
2459592.214846 &  1155.71   &   62.52  &  1217.54  &   199.04 & 1800 & PARAS\\
2459619.190349 &   791.20    & 52.30    &   -92.82  &   249.29 & 1800 & PARAS\\
2459619.224340 &  744.52    &    82.10  &   413.35   &  278.35 & 1800 & PARAS\\
2459619.269151 &    589.99    & 83.52   &   236.22  &   255.28 & 1800 & PARAS\\
2459647.154375 &  91.81     &   58.45   &   -22.06  &   128.41 & 1800 & PARAS\\
2459647.190552 &  -18.86    &   54.08   &  -331.87   &   91.57 & 1800 & PARAS\\
2459648.118842 &  584.50    & 84.86   &  -1495.12   &   285.70  & 1800 & PARAS\\ 
2459650.113942 &   1211.70  &  66.83    & -1008.67   &  149.20 & 1800 & PARAS\\
2459650.191793 &  1082.77    &  86.56   &  -1048.42   &  206.49 & 1800 & PARAS\\
2459651.119375 &  160.29    &    62.99 &   -451.66  &   147.03 & 1800 & PARAS\\
2459651.150425 &  251.80    &  58.78    &  -991.27  &   247.10 & 1800 & PARAS\\
2459673.160786 &  -35.14     &  82.03     &  259.99  &  224.52  & 1800 & PARAS\\
2459676.145032 & -172.45      & 79.06   &  -2088.29   &  231.24 & 1800 & PARAS\\
2459678.118658&  1374.26     &  99.56      & -745.09  &   421.19 & 1800 & PARAS\\
2459881.366954 & 1586.09   & 59.53      &   -1033.86  &    151.34 & 1800 & PARAS\\
2459881.390763 &  1655.28   &   70.41  &  -1160.88  &   206.22 & 1800 & PARAS\\
2459882.340429 &   831.30   & 56.21    &    -414.69  &   100.21 & 1800 & PARAS\\
2459882.363486 &  838.82    &    52.98  &   248.93   &  439.60 & 1800 & PARAS\\
2459882.498600 &   390.78    & 50.84   &   -417.80  &   230.02  & 1800 & PARAS\\
2459883.322491 &   16.65    &   48.72 &  -1154.95   &  86.43 & 1800 & PARAS\\
2459883.346278 &  -116.47     &   44.56   &   -393.03  &   170.19  & 1800 & PARAS\\
2459884.297863 &  -287.89    &   56.43   &  -2149.04   &   106.84 & 1800 & PARAS\\
2459884.321626 &   -328.58  &  48.01    &  -1984.25   &   99.19 & 1800 & PARAS\\
2459885.323514 &  -226.52    &  96.69   &  -4272.63   &  284.06 & 1800 & PARAS\\
2459886.321280 &  25.85    &    55.50 &   -3578.19  &   330.80 & 1800 & PARAS\\
2459886.418752 &  1.46    &  68.72    &  2356.22  &   159.67 & 1800 & PARAS \\
2459521.904947 & -509  &  130  &--&--& 90 & TRES\\
2459525.893170 & 1012  &  69   &--&--& 180 & TRES\\
2459526.860655 & 1436  &  54   &--& --& 450 & TRES\\
2459604.831871 & 234   &  99   &--&--& 270 & TRES\\
2459819.997097 & -552  &  46   &--&--& 360 & TRES\\
2459820.998886 & -396  &  58   &--&--& 180 & TRES\\
2459824.001461 & 1193  &  96   &--&--& 195 & TRES\\
2459824.981289 & -130  &  50   &--&--& 400 & TRES\\
2459829.013338 & 0.00  &  78   &--&--& 720 & TRES\\
2459830.019859 & 1210  &  116  &--&--& 300 & TRES\\
2459836.992174 & 470   &  59   &--&--& 210 & TRES\\
2459837.976241 & 1474  &  95   &--&--& 180 & TRES\\
2459839.014168 & 540   &  78   &--&--& 240 & TRES \\
	\hline
\end{longtable}
 
\end{center}

\newpage
\onecolumn
\begin{longtable}{lcccc}
\caption{\textcolor{black}{}Priors along with Median values and 68\% confidence interval for TOI-4603 from EXOFASTv2. The $\mathcal{N}$ and $\mathcal{U}$ represent the Gaussian and the Uniform priors, respectively.}
\label{result_exofast}\\
\hline
\noalign{\smallskip}
{Parameter} & {Units} & {Adopted Priors} & {Values}\\
\noalign{\smallskip}
\hline
\noalign{\smallskip}
\multicolumn{2}{l}{Stellar Parameters:}&\smallskip\\
~~~~$M_*$\dotfill &Mass (\(M_\odot\))\dotfill &--&$1.765\pm0.061$\\
~~~~$R_*$\dotfill &Radius (\(R_\odot\))\dotfill &--&$2.738^{+0.048}_{-0.050}$\\
~~~~$L_*$\dotfill &Luminosity (\(L_\odot\))\dotfill &--&$10.40^{+0.65}_{-0.62}$\\
~~~~$\rho_*$\dotfill &Density (cgs)\dotfill &--&$0.1211^{+0.0077}_{-0.0071}$\\
~~~~$\log{g}$\dotfill &Surface gravity (cgs)\dotfill &--&$3.810^{+0.021}_{-0.020}$\\
~~~~$T_{\rm eff}$\dotfill &Effective Temperature (K)\dotfill &$\mathcal{N}$(6169, 128)&$6264^{+95}_{-94}$\\
~~~~$[{\rm Fe/H}]$\dotfill &Metallicity (dex)\dotfill &--&$0.342^{+0.039}_{-0.040}$\\

~~~~$Age$\dotfill &Age (Gyr)\dotfill &--&$1.64^{+0.30}_{-0.24}$\\
~~~~$EEP$\dotfill &Equal Evolutionary Point \dotfill &--&$395.7^{+10.}_{-9.2}$\\
~~~~$A_V$\dotfill &V-band extinction (mag)\dotfill &$\mathcal{U}$(0, 1.5965)&$0.272^{+0.089}_{-0.090}$\\
~~~~$\sigma_{SED}$\dotfill &SED photometry error scaling \dotfill &--&$3.64^{+0.95}_{-0.66}$\\
~~~~\textcolor{black}{$v \sin{i}$}\dotfill &\textcolor{black}{Projected Rotational Velocity (km s$^{-1}$)}\dotfill&-- &\textcolor{black}{$23.18\pm0.37$}\\
~~~~$\varpi$\dotfill &Parallax (mas)\dotfill &$\mathcal{N}$(4.4613, 0.01947)&$4.462\pm0.020$\\
~~~~$d$\dotfill &Distance (pc)\dotfill &--&$224.12\pm0.99$\\
~~~~$\dot{\gamma}$\dotfill &RV slope (m/s/day)\dotfill& -- &$-0.14\pm0.18$\\
\smallskip\\\multicolumn{2}{l}{Planetary Parameters:}&&b\smallskip\\
~~~~$P$\dotfill &Period (days)\dotfill&-- &$7.24599^{+0.00022}_{-0.00021}$\\
~~~~$R_P$\dotfill &Radius (\rj)\dotfill&-- &$1.042^{+0.038}_{-0.035}$\\
~~~~$T_C$\dotfill &Time of conjunction (\bjdtdb)\dotfill &--&$2459549.1260\pm0.0014$\\
~~~~$a$\dotfill &Semi-major axis (AU)\dotfill &--&$0.0888\pm0.0010$\\
~~~~$i$\dotfill &Inclination (Degrees)\dotfill &--&$80.21^{+0.39}_{-0.41}$\\
~~~~$e$\dotfill &Eccentricity \dotfill &--&$0.325\pm0.020$\\
~~~~$\omega_*$\dotfill &Argument of Periastron (Degrees)\dotfill &--&$20.4^{+4.6}_{-4.7}$\\
~~~~$T_{eq}$\dotfill &Equilibrium temperature (K)\dotfill &--&$1677\pm24$\\
~~~~$M_P$\dotfill &Mass (\mj)\dotfill&-- &$12.89^{+0.58}_{-0.57}$\\
~~~~$K$\dotfill &RV semi-amplitude (m/s)\dotfill &--&$962^{+37}_{-35}$\\
~~~~$logK$\dotfill &Log of RV semi-amplitude \dotfill &--&$2.983\pm0.016$\\
~~~~$R_P/R_*$\dotfill &Radius of planet in stellar radii \dotfill &--&$0.0391^{+0.0012}_{-0.0010}$\\
~~~~$a/R_*$\dotfill &Semi-major axis in stellar radii \dotfill &--&$6.97\pm0.14$\\
~~~~$\delta$\dotfill &Transit depth (fraction)\dotfill &--&$0.001528^{+0.000091}_{-0.000079}$\\
~~~~$Depth$\dotfill &Flux decrement at mid-transit \dotfill &--&$0.001528^{+0.000090}_{-0.000079}$\\
~~~~$T_{14}$\dotfill &Total transit duration (days)\dotfill&-- &$0.1189\pm0.0022$\\
~~~~$b$\dotfill &Transit Impact parameter \dotfill&-- &$0.9521^{+0.0044}_{-0.0049}$\\
~~~~$\rho_P$\dotfill &Density (cgs)\dotfill&-- &$14.1^{+1.7}_{-1.6}$\\
~~~~$logg_P$\dotfill &Surface gravity \dotfill&-- &$4.469^{+0.036}_{-0.037}$\\
~~~~$\fave$\dotfill &Incident Flux (\fluxcgs)\dotfill&-- &$1.622^{+0.097}_{-0.092}$\\
~~~~$T_P$\dotfill &Time of Periastron (\bjdtdb)\dotfill&-- &$2459548.363^{+0.075}_{-0.083}$\\
~~~~$ecos{\omega_*}$\dotfill & \dotfill&-- &$0.303\pm0.019$\\
~~~~$esin{\omega_*}$\dotfill & \dotfill&-- &$0.113\pm0.027$\\
~~~~$M_P\sin i$\dotfill &Minimum mass (\mj)\dotfill&-- &$12.70^{+0.57}_{-0.56}$\\
~~~~$M_P/M_*$\dotfill &Mass ratio \dotfill&-- &$0.00698^{+0.00028}_{-0.00027}$\\
\noalign{\smallskip}
\noalign{\smallskip}\noalign{\smallskip}
\hline
\noalign{\smallskip}
\multicolumn{2}{l}{Wavelength Parameters:}&TESS\smallskip\\
~~~~$u_{1}$\dotfill &linear limb-darkening coeff \dotfill &$0.237\pm0.050$\\
~~~~$u_{2}$\dotfill &quadratic limb-darkening coeff \dotfill &$0.318\pm0.050$\\
\smallskip\\\multicolumn{2}{l}{Telescope Parameters:}&PARAS&TRES\smallskip\\
~~~~$\gamma_{\rm rel}$\dotfill &Relative RV Offset (m/s)\dotfill &$376^{+23}_{-24}$&$147^{+59}_{-56}$\\
~~~~$\sigma_J$\dotfill &RV Jitter (m/s)\dotfill &$95^{+25}_{-20}$&$185^{+68}_{-51}$\\
~~~~$\sigma_J^2$\dotfill &RV Jitter Variance \dotfill &$9100^{+5400}_{-3400}$&$35000^{+30000}_{-16000}$\\
\smallskip\\\multicolumn{2}{l}{Transit Parameters:}& TESS (TESS)\smallskip\\
~~~~$\sigma^{2}$\dotfill &Added Variance \dotfill &$0.0000000151\pm0.0000000018$\\
~~~~$F_0$\dotfill &Baseline flux \dotfill &$1.0000094\pm0.0000024$\\
\noalign{\smallskip}
\hline
\noalign{\smallskip}
\end{longtable}

\twocolumn
\section{Figures}

\begin{figure}[h!]
\centering
	\includegraphics[width=\columnwidth]{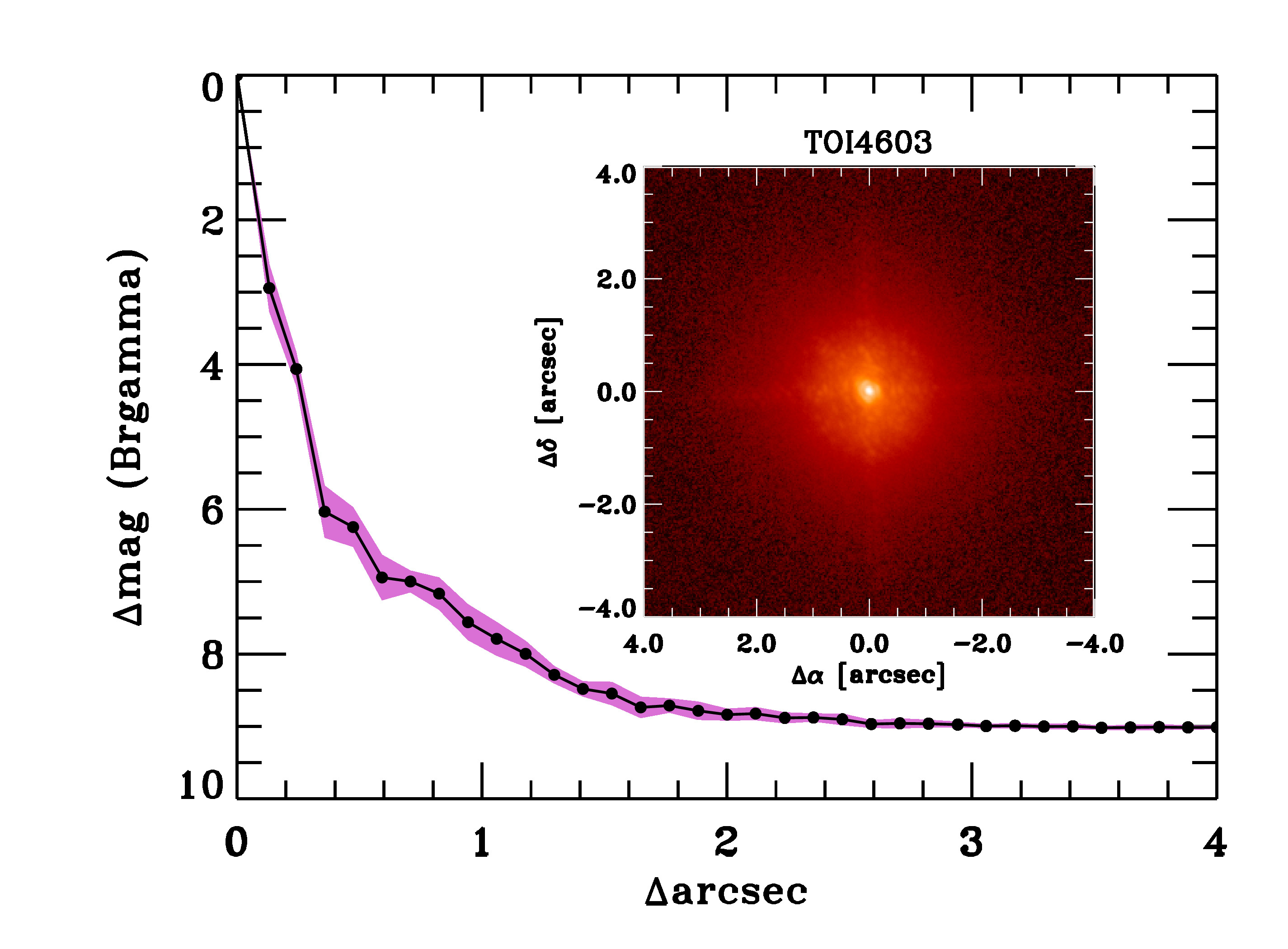}
    \caption{Palomar NIR AO imaging and sensitivity curves for TOI-4603 taken in the Br$\gamma$ filter. The images were taken in good seeing conditions, and we reach a contrast of $\sim 7$ magnitudes fainter than the host star within 0.\arcsec5. {\it Inset:} Image of the central portion of the data, centered on the star.}
    \label{fig:palomar_ao}
\end{figure}

\begin{figure}[h!]
    \centering
	\includegraphics[width=\columnwidth]{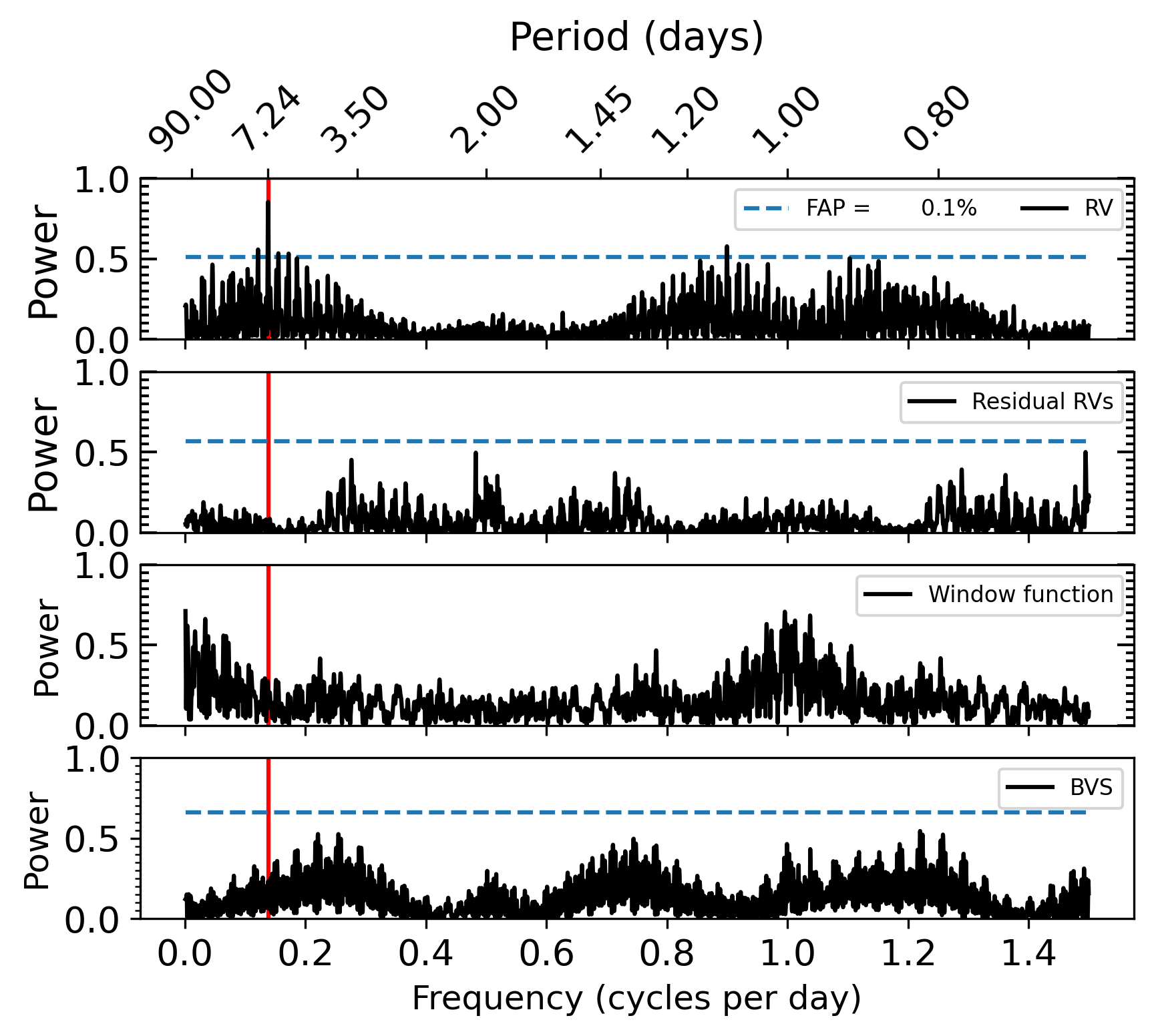}
    \caption {The GLS periodogram for the RVs, residual RVs, window function, and bisector slope of TOI-4603 is shown in panel \texttt{1}, \texttt{2}, \texttt{3}, and \texttt{4} (upper to lower) respectively. The primary peak is seen at a period $\approx 7.24$ days (red vertical line), consistent with the orbital period of the planetary candidate obtained from photometry. The FAP levels (dashed lines) of 0.1$\%$ for all the periodograms are shown in the legends of the panel~\texttt{1}.}
    \label{fig:periodogram}
\end{figure}

\begin{figure}
    \centering
	\includegraphics[width=0.90\columnwidth]{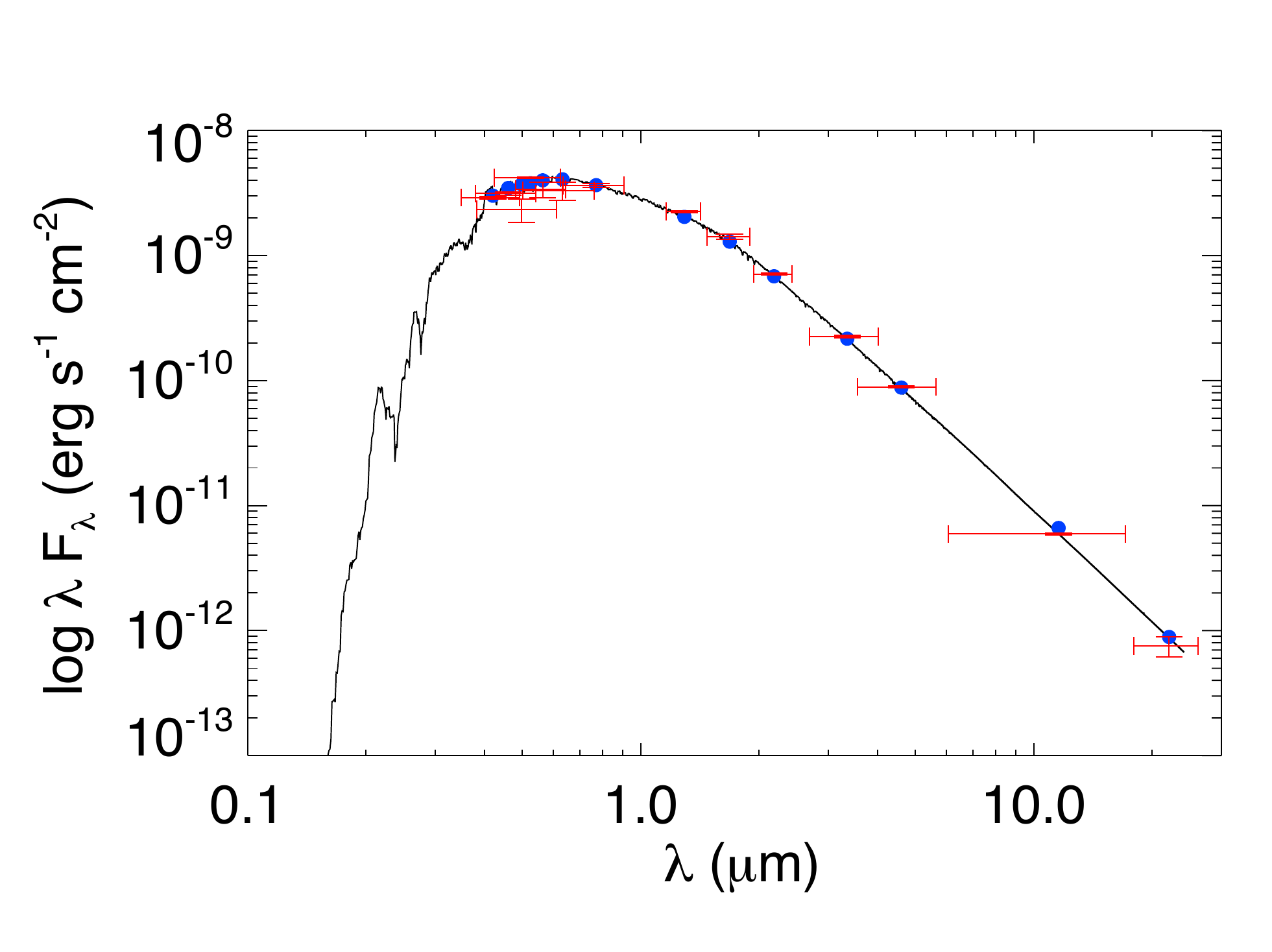}
    \caption{\textcolor{black}{}The spectral energy distribution of TOI-4603.}
    \label{fig:sed}
\end{figure}

\begin{figure}
	\centering
	\includegraphics[width=0.75\columnwidth]{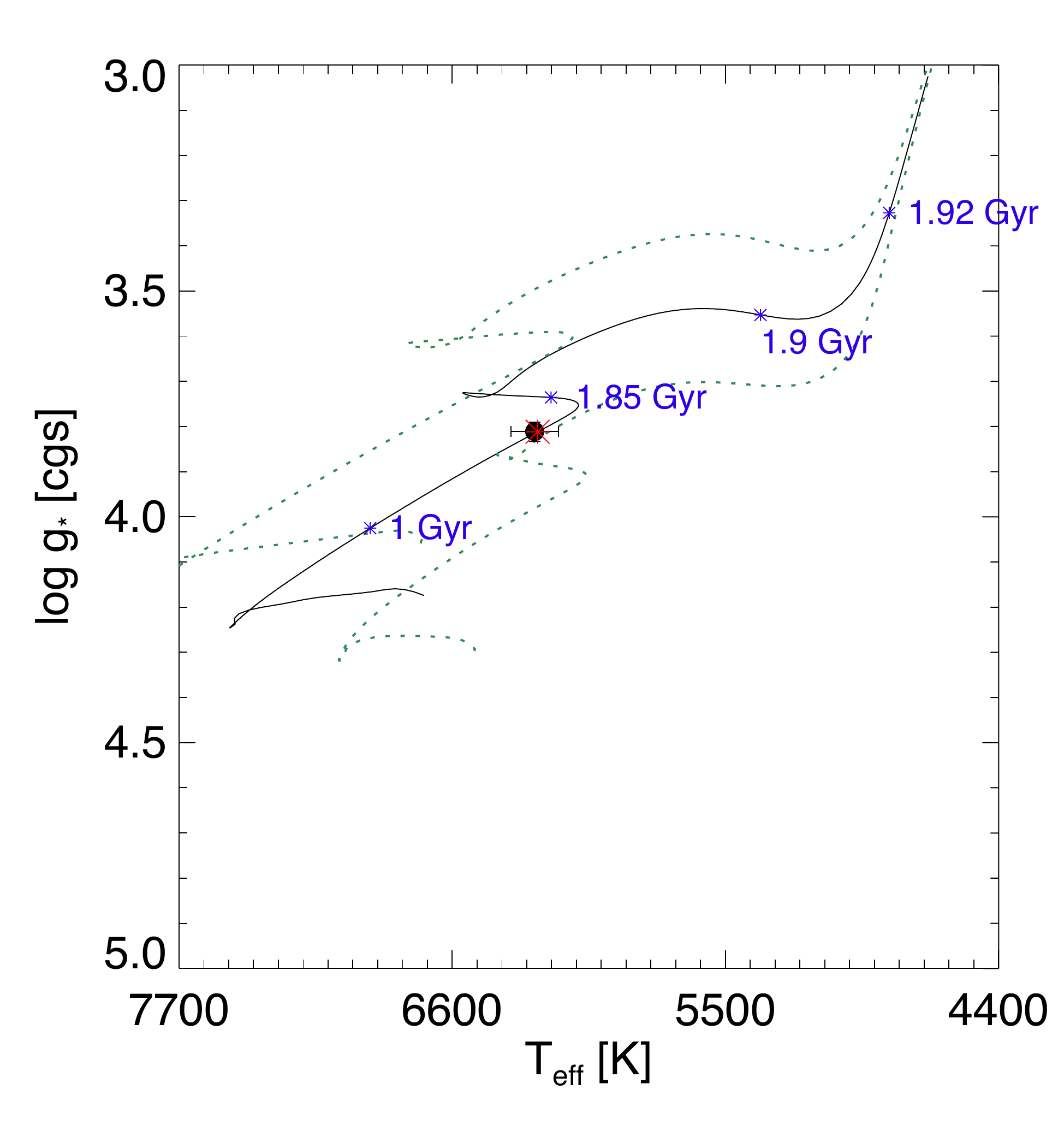}
    \caption{\textcolor{black}{}The most likely MIST evolutionary track from EXOFASTv2 for TOI-4603 represnted by solid black line. The other two green dashed lines show the evolutionary track of 1.58~\(M_\odot\) and 1.95~\(M_\odot\) (for 3-$\sigma$ limits).}
    \label{fig:mist}
\end{figure}

\begin{figure}
\centering

  \includegraphics[width=0.90\columnwidth]{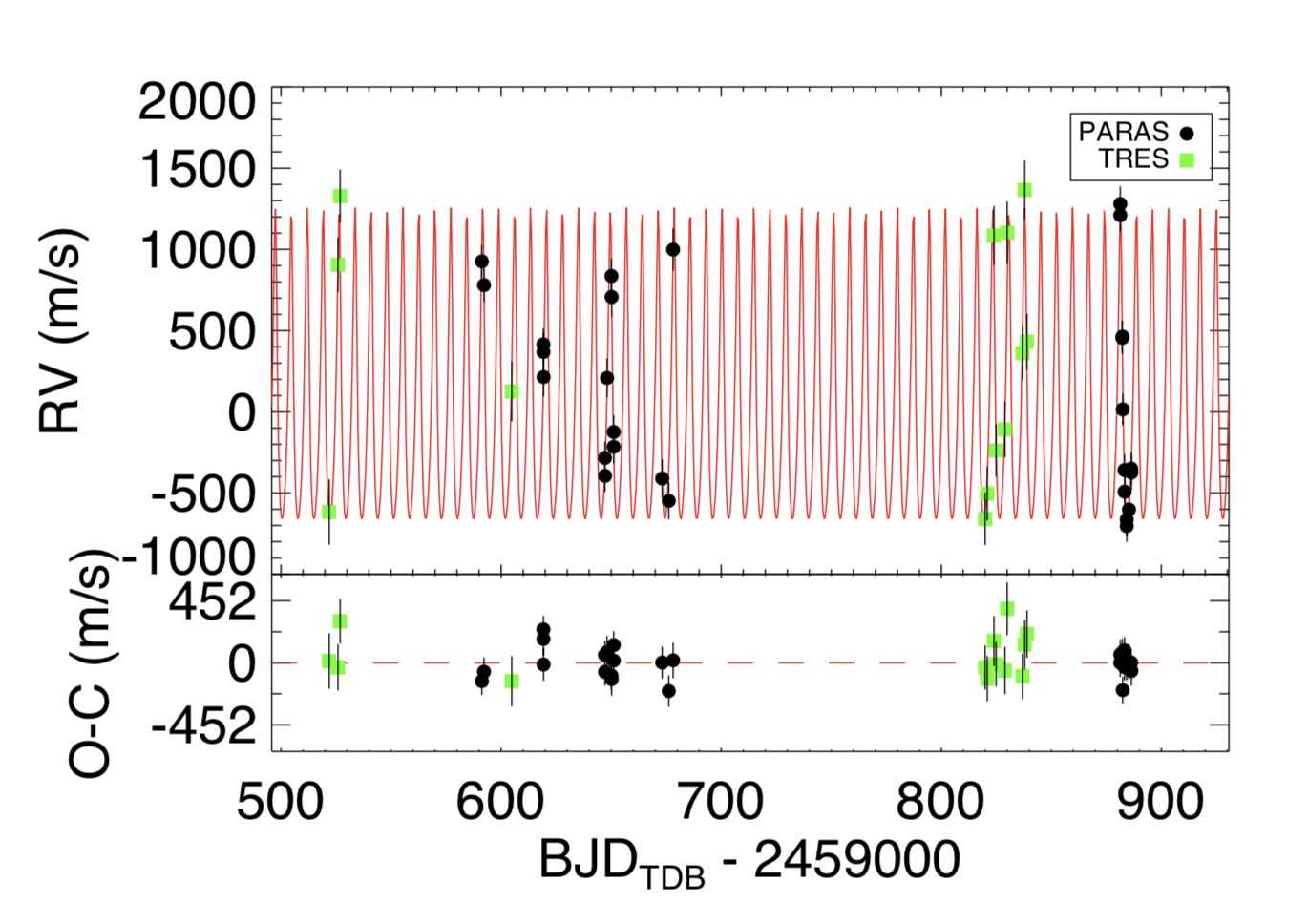}

\caption{\textcolor{black}{}The obtained RVs from PARAS and TRES are plotted with respect to time. The best fit RV model with EXOFASTv2 (see Section~\ref{sec:global}) is represented by the red line, and residuals between the best fit model and the data are shown in the bottom panel. }
    \label{fig:RV_curve}
\end{figure}

\begin{figure*}
\centering
\setkeys{Gin}{width=0.32\linewidth}
\includegraphics{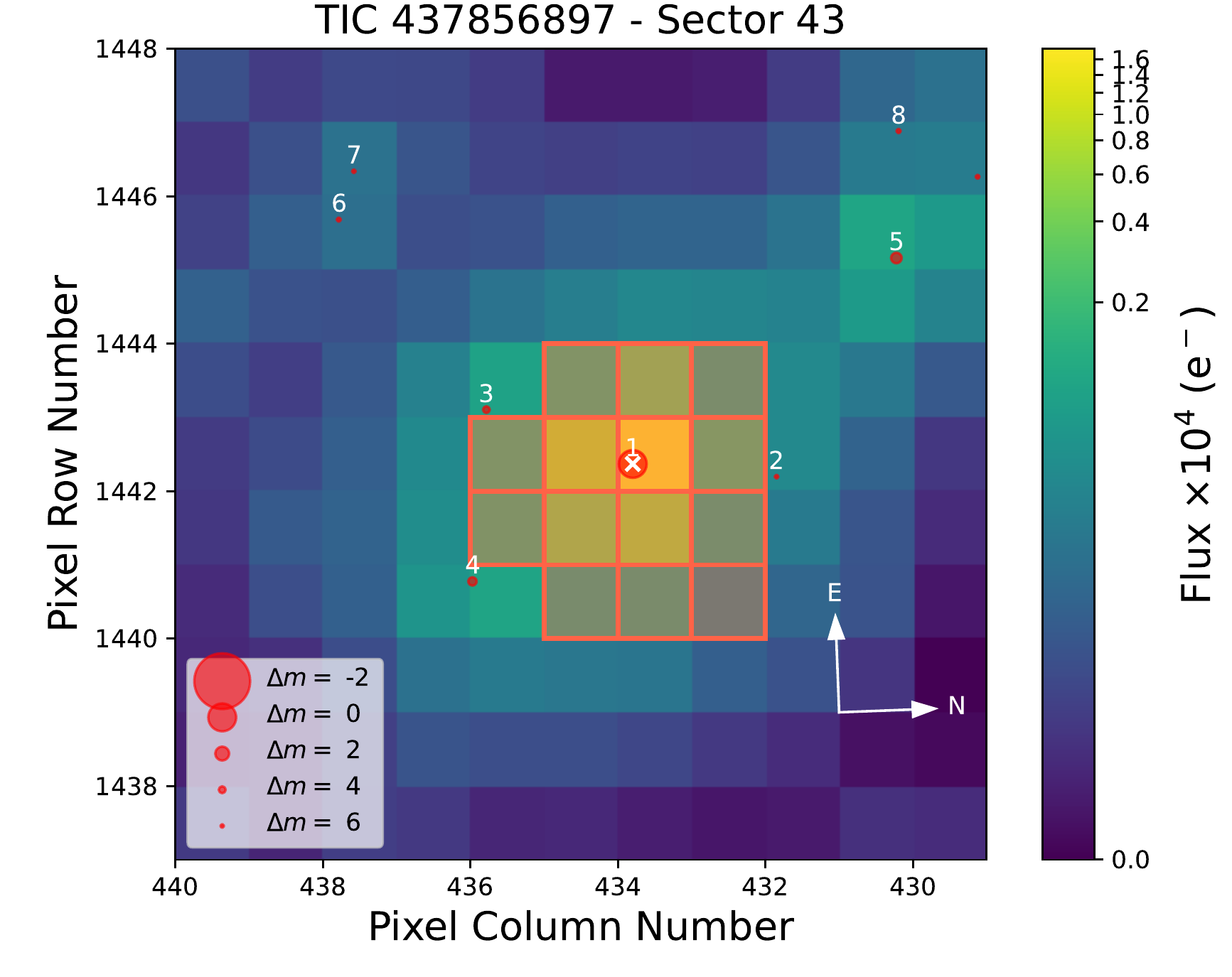}
\hfill
\includegraphics{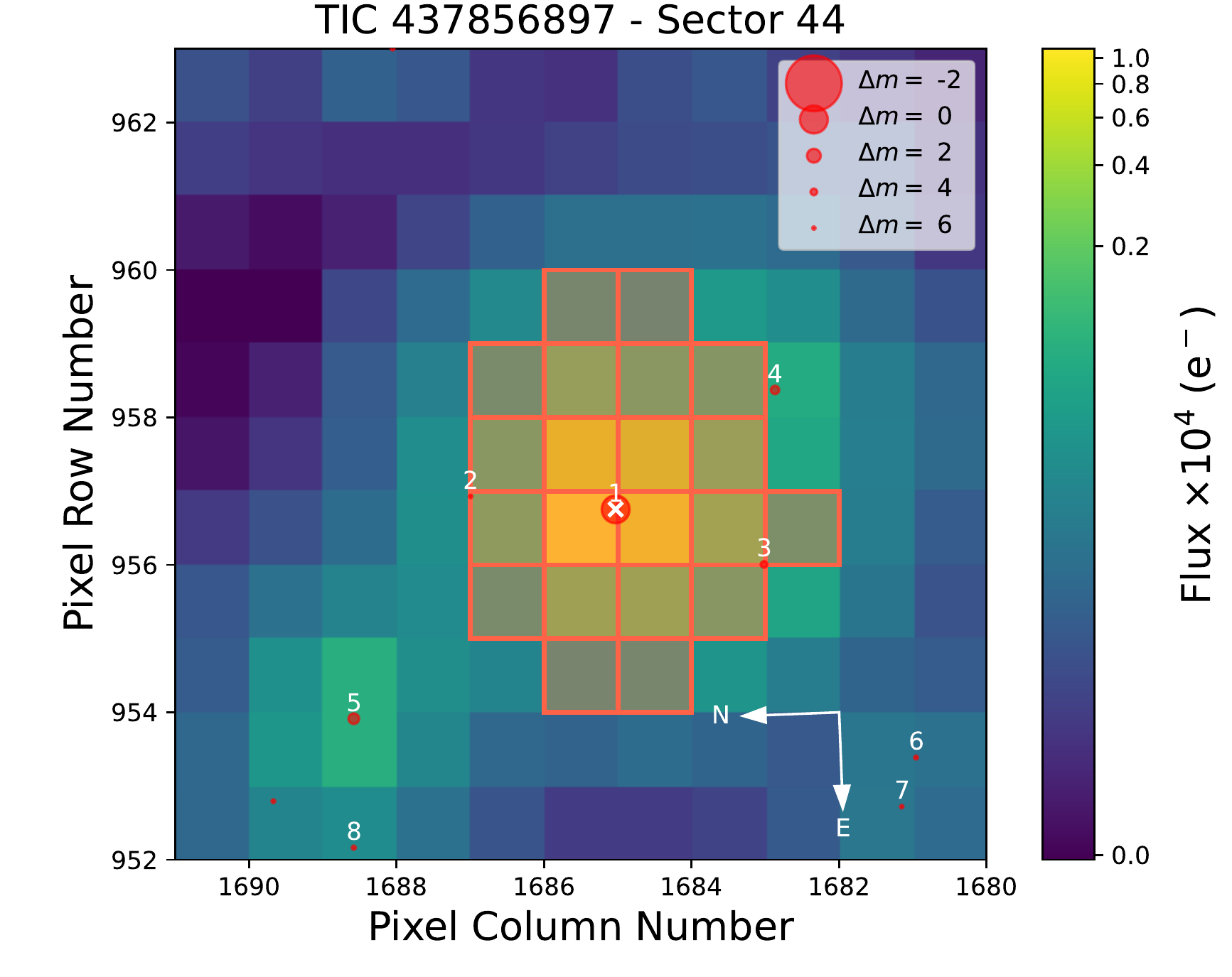}
\hfill
\includegraphics{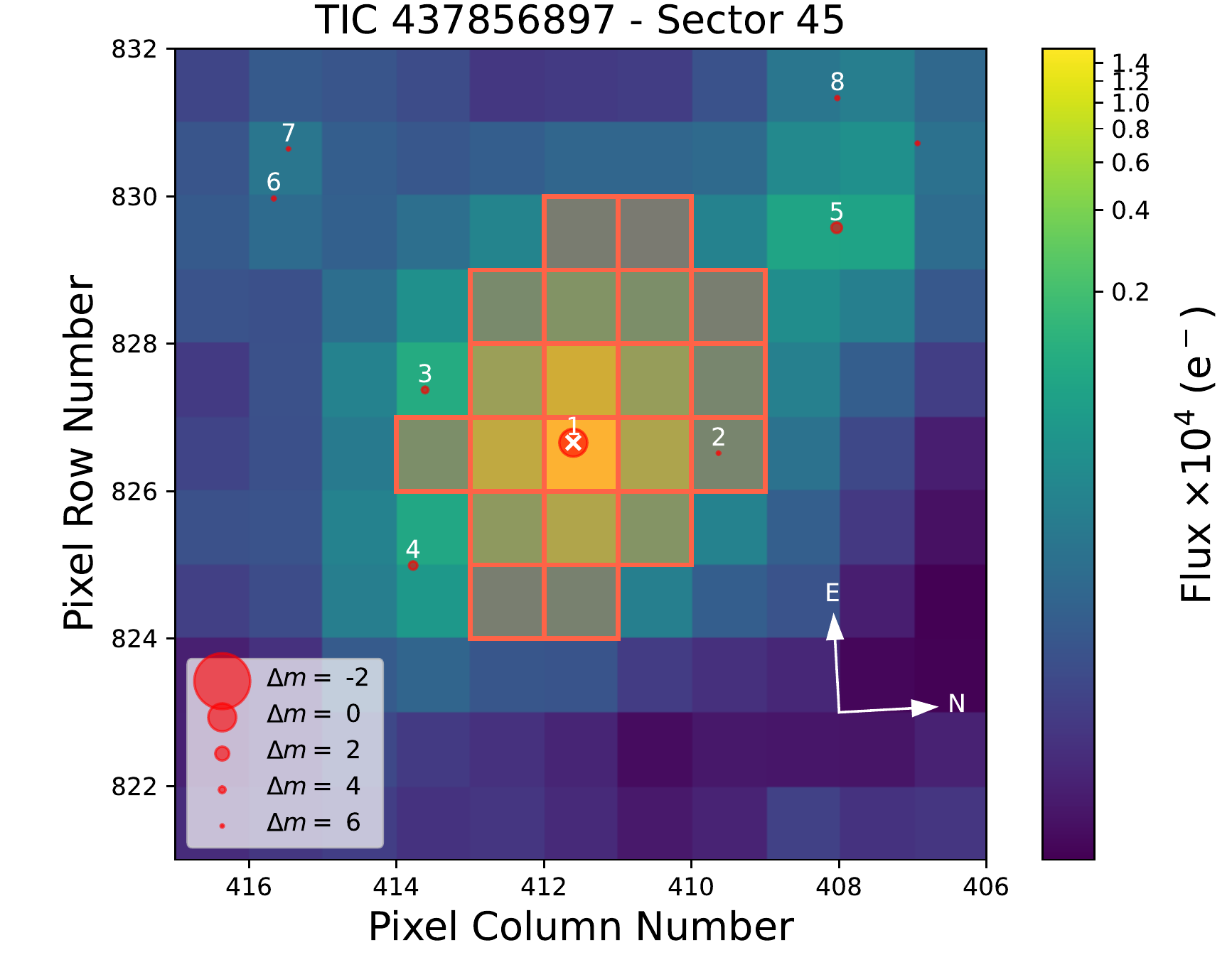}
    \caption{The figure represents the target pixel file for TOI-4603 in sector 43, 44 and 45 generated with \texttt{tpfplotter} \citep{tpfplot}. The squared region is the aperture mask used in the photometry, whereas the size of the individual dot is the magnitude contrast ($\Delta$\textit{m}) from TOI-4603. The position of TOI-4603 is marked with `1'.}
\label{fig:tpfplot}
\end{figure*}

\onecolumn
\newpage
\section{The corner plot showing the covariances for all the fitted parameters for the TOI-4603 global-fit}
\begin{figure}[ht]
\centering
	\includegraphics[width=0.90\columnwidth]{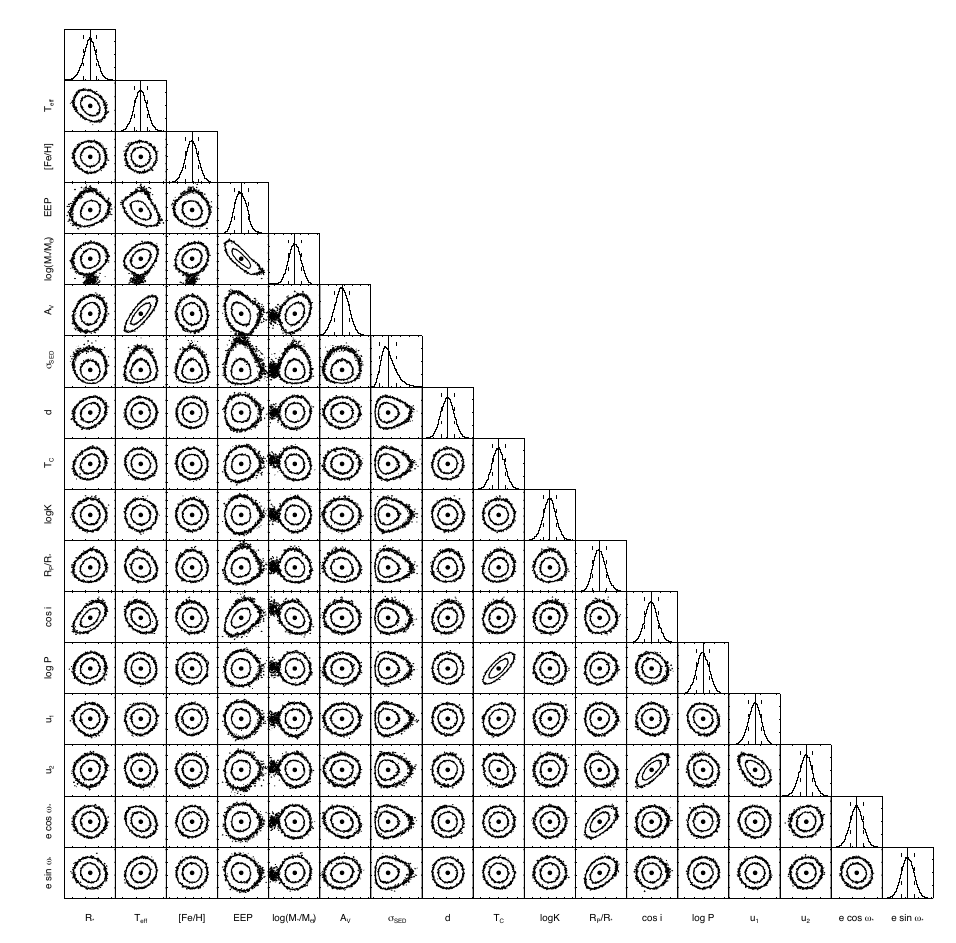}
    \label{fig:corner}
    
\end{figure}
\onecolumn
\newpage
\section{Posterior distribution inferred for the interior modeling of TOI-4603~\lowercase{b}}

\begin{figure}[h!]
    \centering
    \includegraphics[width=0.9\textwidth]{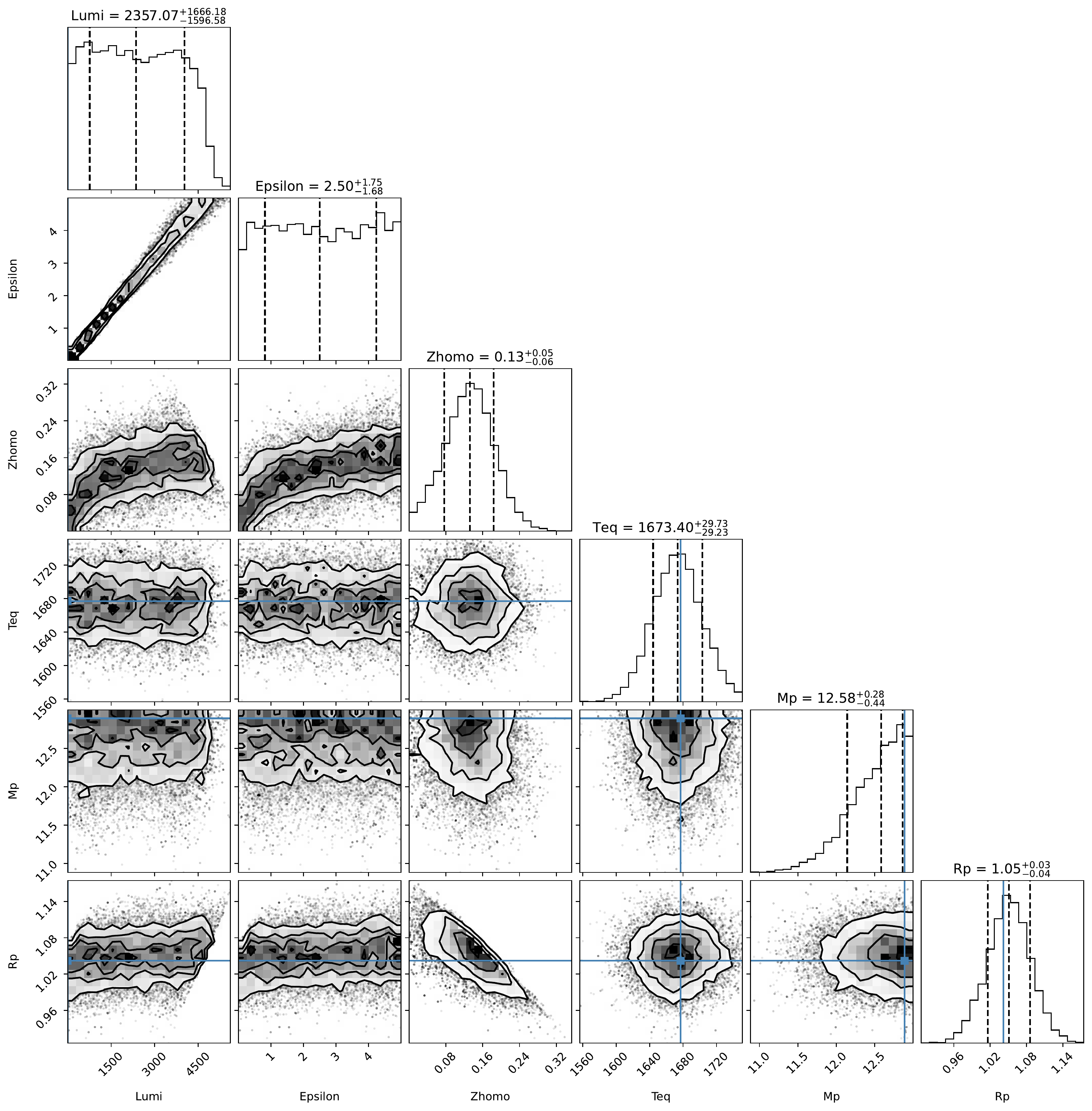} 
\end{figure}

\end{appendix}

\end{document}